\documentclass[useAMS,usenatbib,twocolumn]{mn2e}
\usepackage{graphicx}
\usepackage[english]{babel}
\usepackage{multirow}
\usepackage{lscape}
\usepackage{longtable}
\usepackage{natbib}
\usepackage[usenames,dvipsnames]{color}
\usepackage{float}
\usepackage{url}
\usepackage[T1]{fontenc}

\def\LaTeX{L\kern-.36em\raise.3ex\hbox{a}\kern-.15em
    T\kern-.1667em\lower.7ex\hbox{E}\kern-.125emX}

\begin{document}

\voffset-1.25cm

   \title[LoRCA]{The Low Redshift survey at Calar Alto (LoRCA)}
      
   \author[J. Comparat et al.]
   {
   \parbox{\textwidth}{J. Comparat,$^{1,2,}$\thanks{j.comparat@csic.es}$^,$\thanks{Severo Ochoa IFT Fellow} C.-H. Chuang,$^{1}$$^,$\thanks{MultiDark Fellow} S. Rodr\'iguez-Torres,$^{1,3,2,}$\thanks{Campus de Excelencia Internacional UAM/CSIC Scholar} M. Pellejero-Ibanez,$^{4,5,6}$ F. Prada,$^{1,3,7}$ G. Yepes,$^2$ H. M. Courtois$^8$, G.-B. Zhao,$^{9,10}$ Y. Wang,$^{9,10}$ J. Sanchez,$^7$ C. Maraston,$^{10}$ R. Benton Metcalf,$^{11}$ J. Peiro-Perez$^1$, F. S. Kitaura$^{12}$, E. P\'erez$^7$, R.~M. Gonz{\'a}lez Delgado$^7$.}
%J. Comparat, C.-H. Chuang, S. Rodriguez-Torres, M. Pellejero-Ibanez, F. Prada, G. Yepes, H. M. Courtois, G.-B. Zhao, Y. Wang, J. Sanchez, C. Maraston, R. Benton Metcalf, J. Peiro-Perez, F. S. Kitaura, E. Perez, R. M. Gonzalez Delgado
\vspace*{6pt} \\
$^1$Instituto de F\'{\i}sica Te\'orica, (UAM/CSIC), Universidad Aut\'onoma de Madrid,  Cantoblanco, E-28049 Madrid, Spain \\
$^2$Departamento de Fisica Teorica, Universidad Autonoma de Madrid,  Cantoblanco E-28049, Madrid, Spain\\ 
$^3$Campus of International Excellence UAM+CSIC, Cantoblanco, E-28049 Madrid, Spain \\
$^4$Instituto de Astrof\'{\i}sica de Canarias (IAC), C/V\'{\i}a
L\'{a}ctea s/n, E-38200, La Laguna, Tenerife, Spain\\
$^5$Departamento Astrof\'{i}sica, Universidad de La Laguna (ULL), E-38206 La Laguna, Tenerife, Spain\\ 
$^6$ MultiDark visitor at IFT-UAM/CSIC, Madrid, Spain \\
$^7$ Instituto de Astrof\'{\i}sica de Andaluc\'{\i}a (CSIC), Glorieta de la Astronom\'{\i}a, E-18080 Granada, Spain \\
$^8$University of Lyon, UCB Lyon 1/CNRS/IN2P3, IPN Lyon, France \\ 
$^9$ National Astronomy Observatories, Chinese Academy of Sciences, Beijing, 100012, P.R.China \\
$^{10}$ Institute of Cosmology and Gravitation, Dennis Sciama Building, Burnaby Road, Portsmouth PO1 3FX, UK \\
$^{11}$ Dipartimento di Fisica e Astronomia - Universita di Bologna, via Berti Pichat 6/2, 40127, Bologna, Italy \\
$^{12}$ Leibniz-Institut fur Astrophysik Potsdam (AIP), An der Sternwarte 16, D-14482 Potsdam, Germany
}

\date{\today}

\maketitle
\label{firstpage}

%----------------------abstract----------------------------%   
  \begin{abstract}
The Baryon Acoustic Oscillation (BAO) feature in the power spectrum of galaxies provides a standard ruler to measure the accelerated expansion of the Universe. To extract all available information about dark energy, it is necessary to measure a standard ruler in the local, $z<0.2$, universe where dark energy dominates most the energy density of the Universe. 
Though the volume available in the local universe is limited, it is just big enough to measure accurately the long 100 $h^{-1}$Mpc wave-mode of the BAO. 
Using cosmological $N$-body simulations and approximate methods based on Lagrangian perturbation theory, we construct a suite of a thousand light-cones to evaluate the precision at which one can measure the BAO standard ruler in the local universe.  
We find that using the most massive galaxies on the full sky (34,000 deg$^2$), {\rm i.e.} a $K_\mathrm{2MASS}<14$ magnitude-limited sample, one can measure the BAO scale up to a precision of 4\% ($\sim1.2\%$ using reconstruction). 
We also find that such a survey would help to detect the dynamics of dark energy.
Therefore, we propose a 3-year long observational project, named the Low Redshift survey at Calar Alto (LoRCA), to observe spectroscopically about 200,000 galaxies in the northern sky to contribute to the construction of aforementioned galaxy sample. The suite of light-cones is made available to the public.
   \end{abstract}

%The Baryon Acoustic Oscillation (BAO) feature in the power spectrum of galaxies provides a standard ruler to measure the accelerated expansion of the Universe. To extract all available information about dark energy, it is necessary to measure a standard ruler in the local, z<0.2, universe where dark energy dominates most the energy density of the Universe. 
%Though the volume available in the local universe is limited, it is just big enough to measure accurately the long 100 Mpc/h wave-mode of the BAO. 
%Using cosmological N-body simulations and approximate methods based on Lagrangian perturbation theory, we construct a suite of a thousand light-cones to evaluate the precision at which one can measure the BAO standard ruler in the local universe.  
%We find that using the most massive galaxies on the full sky (34,000 sq. deg.), i.e. a K(2MASS)<14 magnitude-limited sample, one can measure the BAO scale up to a precision of 4\% and 1.2\% using reconstruction). 
%We also find that such a survey would help to detect the dynamics of dark energy.
%Therefore, we propose a 3-year long observational project, named the Low Redshift survey at Calar Alto (LoRCA), to observe spectroscopically about 200,000 galaxies in the northern sky to contribute to the construction of aforementioned galaxy sample. The suite of light-cones is made available to the public.

      
%--------------------------------------------------------------------------------------------------------------
%--------------------------------------------------------------------------------------------------------------
%------------------------------			INTRODUCTION       -----------------------------------
%--------------------------------------------------------------------------------------------------------------
%--------------------------------------------------------------------------------------------------------------
\section{Introduction}
\label{section:introduction}
The combination of current and future dark-energy spectroscopic surveys (e.g. BOSS, eBOSS, DESI, Euclid) will map the distance--redshift relation using the BAO as a standard ruler between redshift 0.2 and 2.5 down to 1-2\% within the next 5 years and 0.3\% within the next 10 years, this using multiple tracers: quiescent and star-forming galaxies, quasars and the Lyman-$\alpha$ forest of quasars. These experiments will accurately determine the expansion history of the universe, and hence, will constrain its matter-energy content. As a complement to these probes, we should exhaust the information available in the local universe that is used to calibrate measurements at higher redshifts.

The total (systematics and statistics) uncertainty on the measurement of the Hubble constant today ($H_0$), combining the supernovae data within a radius of $<$300Mpc (redshift $z<0.08$) whose distances are accurately determined with cepheids, is of $\pm$5\% \citep[see][for a review]{2010ARA&A..48..673F}. This uncertainty should lower to $\pm$2\% with the addition of JWST\footnote{\url{http://www.jwst.nasa.gov/}}.

An alternative way to calibrate the distance -- redshift relation would be to use the large-scale structure of the local universe, in particular the Baryonic Acoustic Oscillation (BAO) feature in the clustering of nearby galaxies to measure both relations: the angular-diameter distance -- redshift relation and the line-of-sight distance -- redshift relation. 
The first estimates with the BAO standard ruler at low redshift were performed on the 6dF galaxy redshift survey, which samples 0.08 Gpc$^3 h^{-3}$ of the local universe with about 75,000 redshifts \citep{2011MNRAS.416.3017B}. They measured the BAO scale at redshift 0.1 with a precision of 4.5\% and $H_0$ with a precision of 4.8\%. 
\citet{Percival_2010} obtained a 2.7\% precision on the BAO scale by analyzing together the 2dFGRS and SDSS data-sets that contain 800,000 galaxies over 9,100 deg$^2$ at redshift $z=0.2$ (corresponding to a volume of 0.2 Gpc$^3 h^{-3}$). Lately, \citet{2015MNRAS.449..835R} reanalyzed SDSS-DR7 using only the most massive galaxies, and measured the BAO scale at redshift 0.15 with a 4\% precision.
The precision on the BAO standard ruler in the local universe is today limited by the volume sampled by the surveys 6dF, 2dF and SDSS \citep{2009MNRAS.399..683J,2001MNRAS.328.1039C,2014ApJS..211...17A}. 
Now that the full-sky has been imaged by 2MASS and WISE \citep{2006AJ....131.1163S,2010AJ....140.1868W}, there is a possibility of increasing the area (and thus the volume) observed by nearby galaxy spectroscopic surveys onto the full-sky limit. 
The BAO method, being subject to very low-level of systematic errors (e.g. see \citealt{2014MNRAS.441...24A} results for the BOSS survey), it is a great way of making precise measurements of the local geometry of the Universe. Moreover a BAO measurement is orthogonal to the SNIa measurements in the standard parameter space of flat $\Lambda$CDM models. Thus their combination will be very efficient in constraining the local distance -- redshift relation, and hence the cosmological parameters \citep[see][]{2013PhR...530...87W}.

For the BAO probe in a limited volume, provided that  a dense enough tracer allows overcoming the shot noise, the most suited tracer of the underlying dark matter density field is the one that maximizes the clustering amplitude (also called the bias). As the bias correlates with the stellar mass \citep{2011ApJ...736...59Z,2013A&A...557A..17M}, a sample that is closest to a mass-selected sample would meet this criterion. The most massive galaxies of the local universe are early-type galaxies and they have their bolometric luminosity dominated by the near-infrared light (1 to 5 microns) emitted by their evolved stellar population. This wavelength range is sampled by the 2MASS $j$, $h$, $k$ bands (1.24, 1.66, 2.16 $\mu$m) and the WISE $w1$, $w2$ bands (3.35, 4.46 $\mu$m). 
All those bands correlate with the stellar mass and are suited to select massive galaxies in the low redshift universe. 
The magnitude limit of the 2MASS surveys is shallower than that of WISE; $k<$14 compared to $w1<$17 (Vega magnitudes). But, the 2MASS $k$-band being the most studied of all those bands for selecting galaxies, we choose the $k$-band to construct our fiducial galaxy sample; see section \ref{sec:galaxy}.

Then, we investigate the limits in precision of a full-sky BAO measurement at low redshift by analyzing a series of mock catalogs of the local universe built using latest $N$-body BigMultiDark Planck simulations \citep{2014arXiv1411.4001K}; see section \ref{sec:mock:catalog}. 
Next, we use EZmocks \citep{2015MNRAS.446.2621C}
to construct a reliable covariance matrix which will provide robust measurements of the uncertainties. We prove that the BAO measurement accuracy can reach 3.8\% with a full sky sample, 6.8\% with half of the sky and 12\% with a quarter of the sky; see section \ref{sec:BAOmeasurement}.

In the section \ref{subsec:redshiftERR}, we discuss the redshift error one can tolerate to obtain such a BAO measurement and show that current photometric redshift estimations are not precise enough. It is therefore timely to obtain spectroscopic redshifts of the full sky to extract all the cosmological information available.

To this aim, in section \ref{sec:LORCA}, we propose a new galaxy spectroscopic redshift survey: the LoRCA survey\footnote{\url{http://lorca-survey.ft.uam.es}}. LoRCA will observe spectra for about 0.2 million galaxies of the north galactic cap that were not previously observed by SDSS with the best suited northern facility: the Schmidt telescope at the Calar Alto observatory. The LoRCA survey would be highly complementary to the Australian TAIPAN survey\footnote{\url{http://www.taipan-survey.org}}, which aims to observe of order of a million low-redshift galaxies in the south galactic cap, starting in 2016.

Together, these surveys will enable the most detailed three-dimensional map yet of the local Universe and achieve the most precise measurement of the local BAO scale.% and constrain dynamical dark energy models; see section \ref{sec:DDE}.
Finally, in section \ref{sec:AddCosmoProbes}, we discuss additional cosmological probes with LoRCA: peculiar velocities, the galaxy mass function and strong lensing.

Throughout the paper, we assume Planck cosmological parameters $\Omega_\Lambda=0.693$, $\Omega_m=0.307$ \citep{2013arXiv1303.5076P}.

%%%%%%%%%%%%%%%%%%%%%%%%%%%%%%%%%
% 		\section{Galaxy samples}
%%%%%%%%%%%%%%%%%%%%%%%%%%%%%%%%%
\section{Galaxy sample}
\label{sec:galaxy}

\begin{figure*}
\begin{center}
\includegraphics[width=0.4\textwidth]{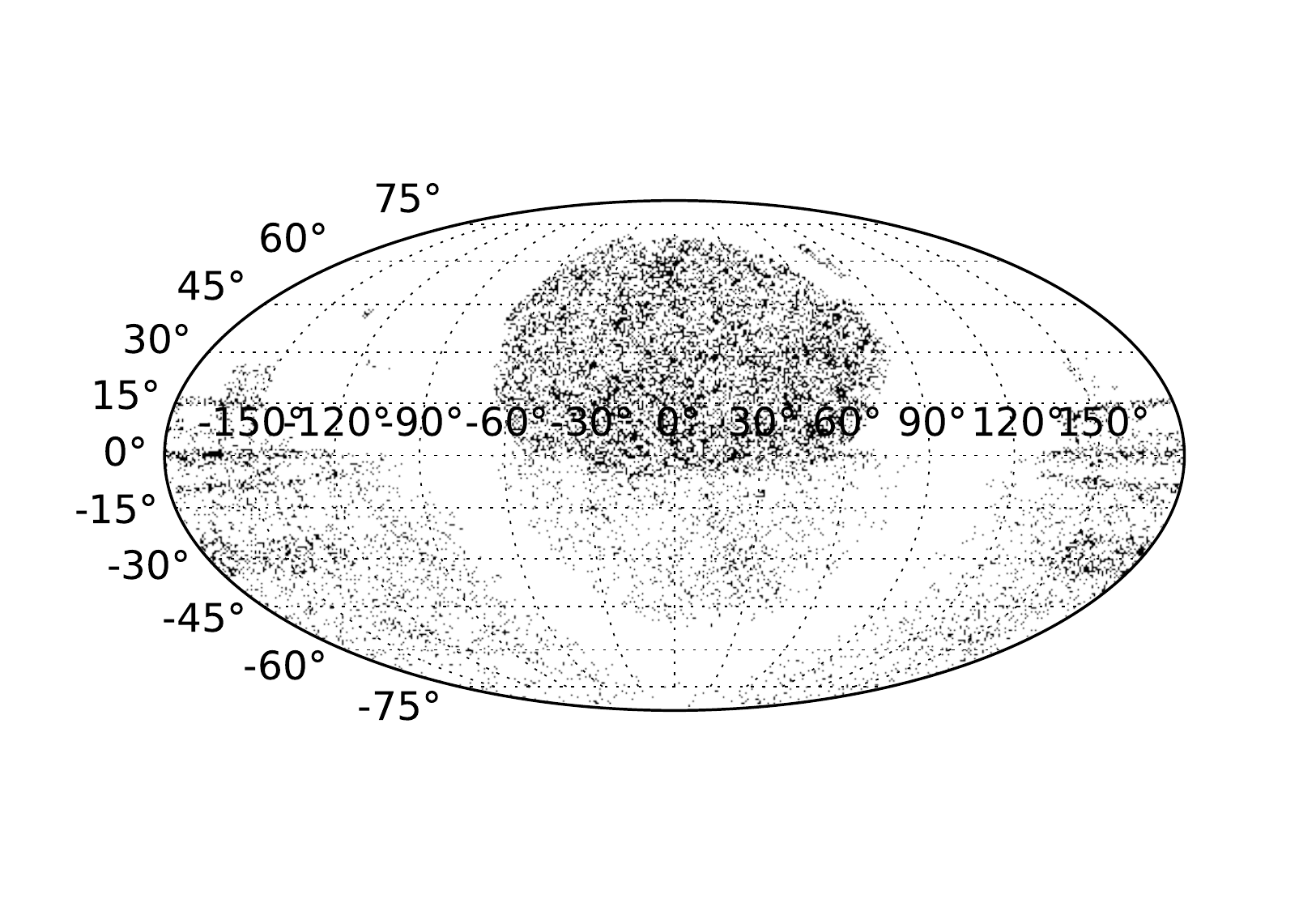}
\includegraphics[width=0.4\textwidth]{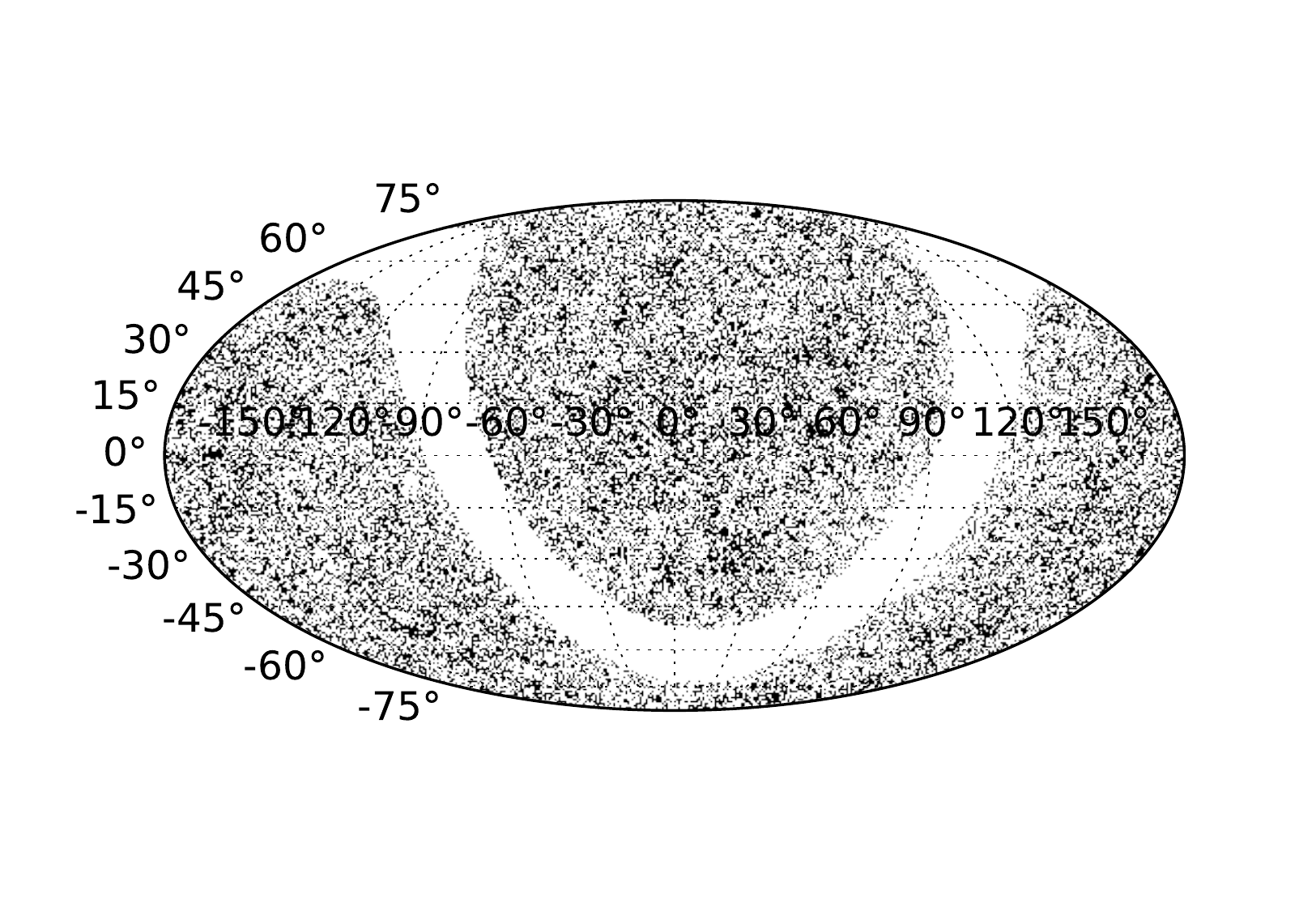}
\caption{DEC vs. RA in degrees (J2000) of the 2MASS $k<14$, $|b_{lat}|>10$ galaxy sample that have been confirmed spectroscopically (left panel) and the map that LoRCA and TAIPAN will provide (right panel).}
\label{kselection:radec}
\end{center}
\end{figure*}

The ideal galaxy sample that would maximize the clustering amplitude is a mass-selected sample. At $z<0.3$, the magnitude that correlates best with the stellar mass is the $k$-band \citep{1998MNRAS.297L..23K}.

We therefore select a $k<14$ magnitude limited sample from the 2MASS full sky survey extended objects catalog \citep{2006AJ....131.1163S}. This catalog contains all sources with a signal-to-noise ratio greater than 7 in at least one of the J,H,K bands, and some morphology criterion probing its extent\footnote{\url{http://www.ipac.caltech.edu/2mass/releases/allsky/doc/explsup.html}}. 
The extended source catalog contains less than 1\% of artifacts and a very small population of stars, between 5 and 10\% depending on Milky Way stellar density. 
Target selection for LoRCA will be refined in a second phase using 2MASS point source catalogs and WISE ultimate photometric catalogs; see discussions in \citet{2014arXiv1401.0156K,2015arXiv150803046R}. 
We keep extended objects with a galactic latitude $|g_{lat}|>10$. Therefore, by masking the Milky Way from the extended source catalog, we obtain a galaxy catalog with purity $>90\%$. Note that for galaxies fainter than $k>14$ the extended source catalog is not complete. In total, we have 853,458 objects over 34,089 deg$^2$ (82.6\% of the full-sky), {\it i.e.} a density of 25 galaxies deg$^{-2}$. Note that by masking further the Milky Way $|g_{lat}|>$15 (20) the area diminishes to 30,575 deg$^2$ (27,143).

\begin{figure*}
\begin{center}
\includegraphics[width=0.45\textwidth]{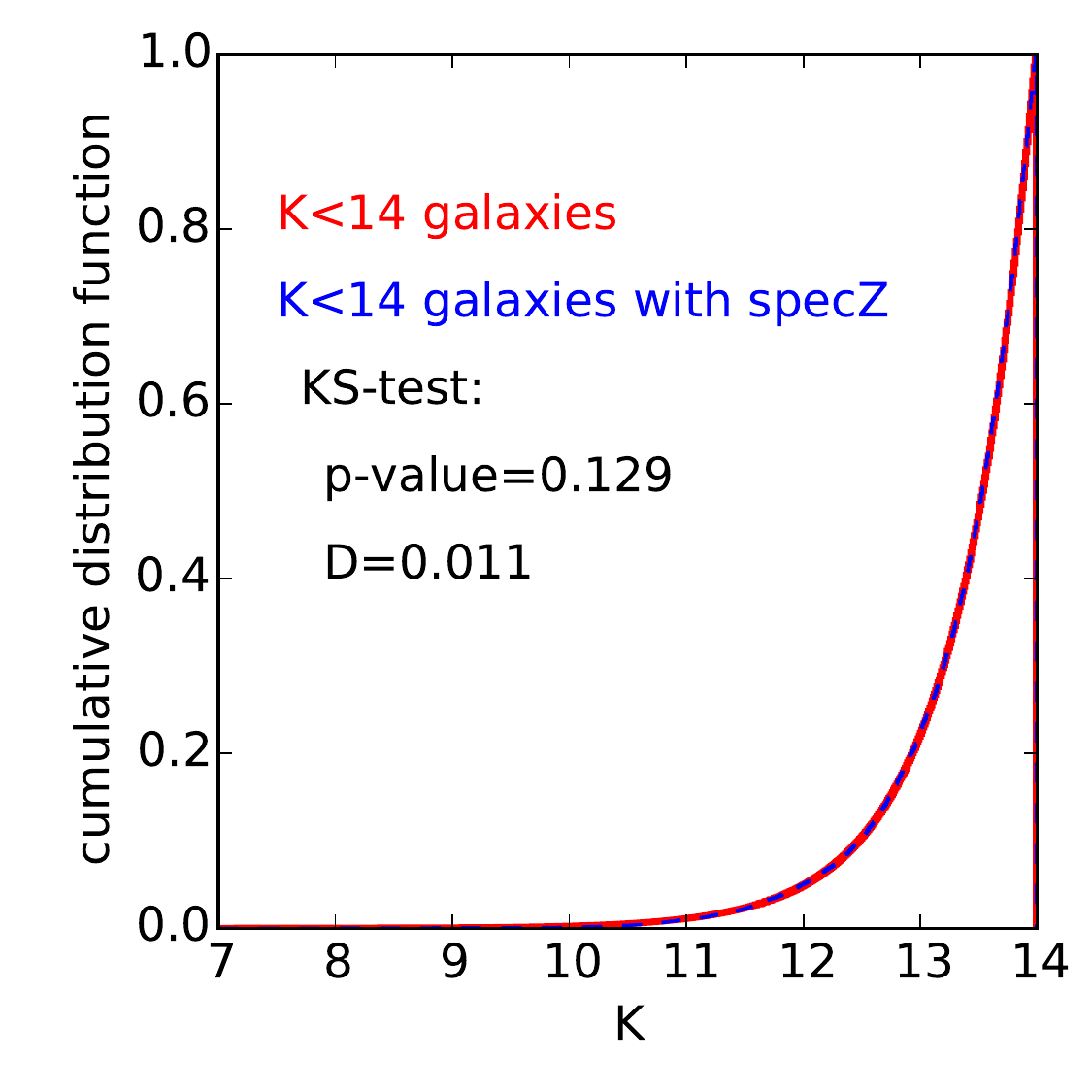}
\includegraphics[width=0.45\textwidth]{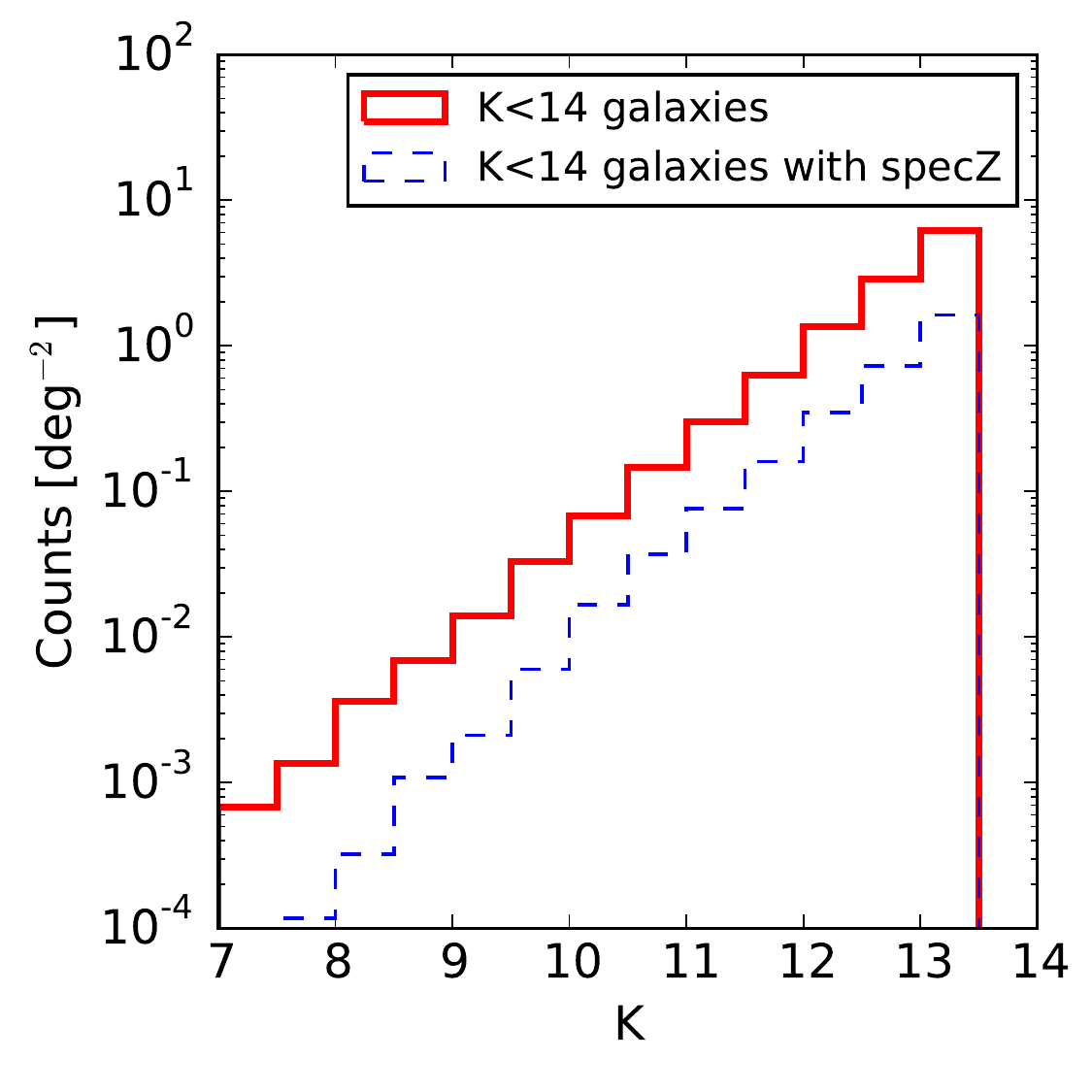}
\caption{Left: $k$-band cumulative distribution function of the complete sample and the SDSS spectroscopic redshift sample. The KS-test result shows that the spectroscopic sample is representative of the entire population. Right: Counts of all and galaxies with spectroscopic redshifts as a function of the K magnitude.}
\label{kselection:kstest}
\end{center}
\end{figure*}

\begin{figure}
\begin{center}
\includegraphics[width=0.4\textwidth]{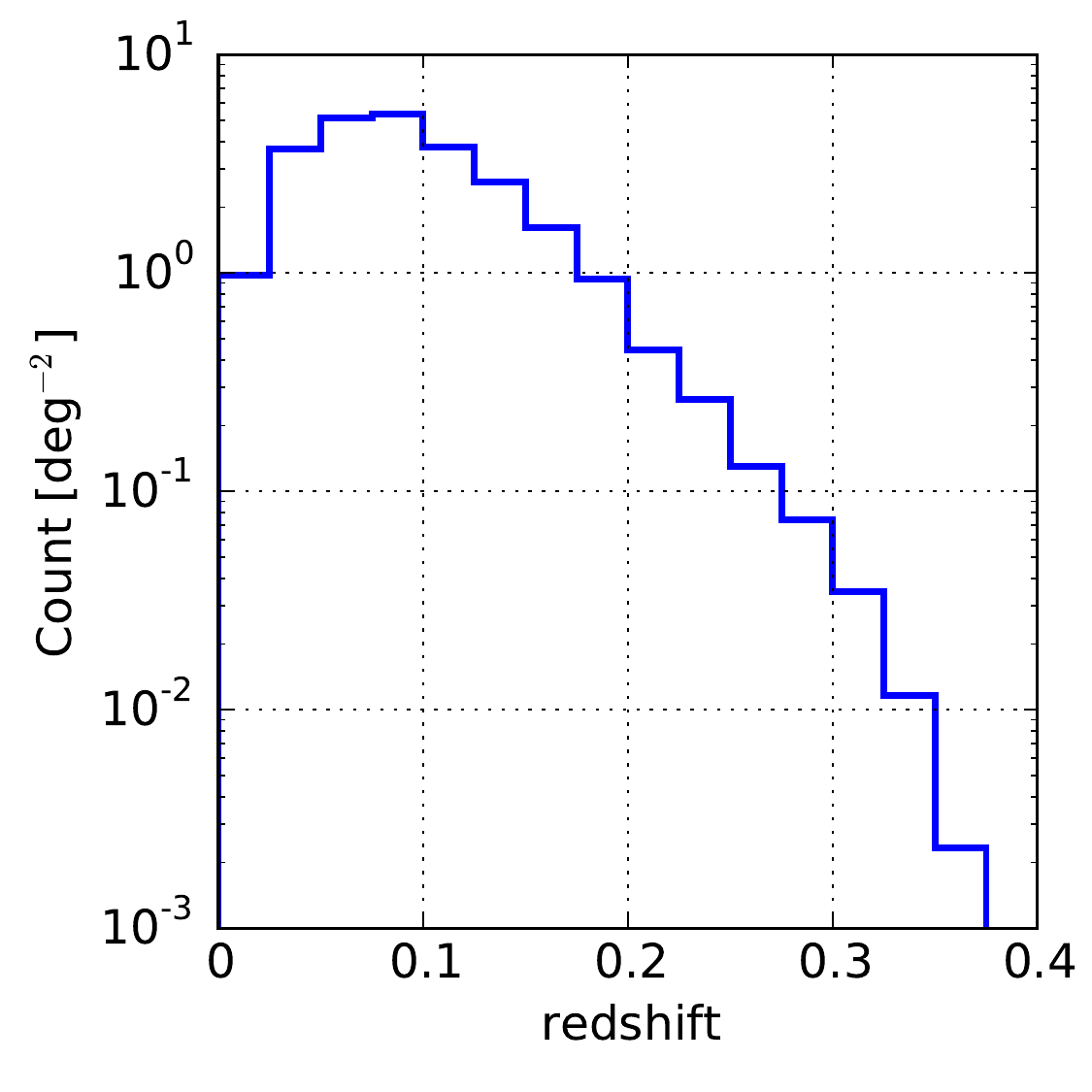}
\caption{Extrapolated galaxy density (in deg$^{-2}$) of the whole $k<14$ 2MASS target sample based on the already observed SDSS spectroscopic sample.}
\label{kselection:nz}
\end{center}
\end{figure}

To obtain a reliable expected redshift distribution of the selected target sample, we combine the following redshift catalogs: the third reference catalog of bright galaxies \citep{1994AJ....108.2128C}, the Sloan Digital Sky Survey DR10 \citep{2014ApJS..211...17A}, the WiggleZ Survey DR1 \citep{Drinkwater_2010}, the 6dFGRS DR3 \citep{2009MNRAS.399..683J}, the GAMA DR2 \citep{2011MNRAS.418.1587T}, the NASA-Sloan Atlas\footnote{\url{http://www.nsatlas.org/}} \citep{2011AJ....142...31B} the 2MASS redshift survey \citep{2012ApJS..199...26H}, and the 2dFGRS \citep{2000ASPC..200...63M}.
In total, about 36.5\% of those targets already have a reliable spectroscopic redshift provided by one of those surveys mentioned above; see Fig. \ref{kselection:radec}. 
A two sample KS-test shows that the available SDSS spectroscopic redshifts constitute a fair sample of the entire population; see left panel in Fig. \ref{kselection:kstest}. Hence, we infer the redshift distribution of the complete sample based on the current spectroscopic observations from SDSS. The right panel of Fig. \ref{kselection:kstest} shows the galaxy counts as a function of k magnitude. 
We rescale the spectroscopic redshift distribution to obtain the same galaxy number density as in the 2MASS photometric selected sample; see Fig. \ref{kselection:nz}. 

\begin{figure}
\begin{center}
\includegraphics[width=0.5\textwidth]{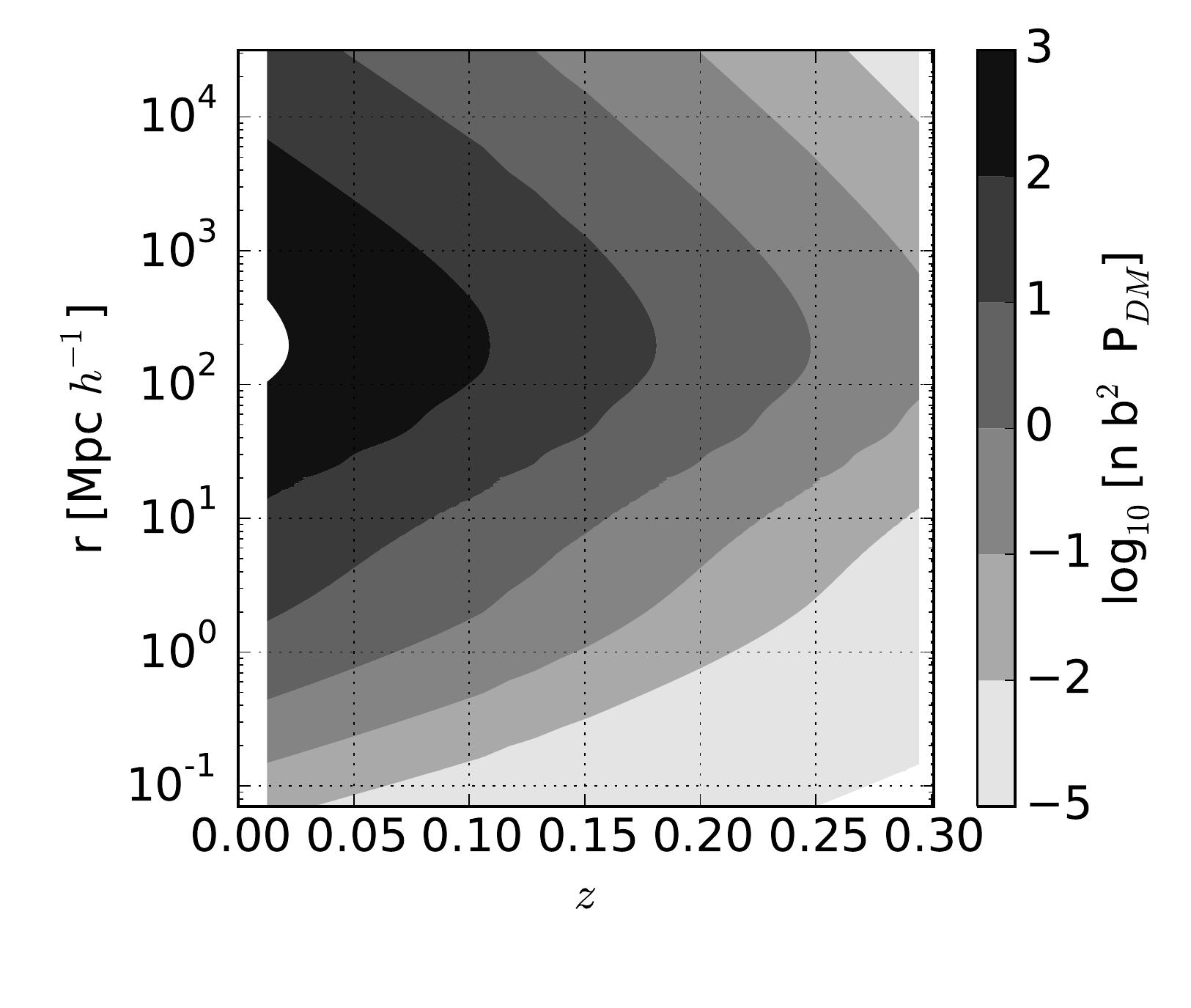}
\caption{ Scale vs. redshift coded with $\rm \log_{10} n(z)\, b^2\, P_{DM}(k,z)$, assuming a bias of b=1.5. The contour labeled 0, corresponding to $n(z)\, b^2\, P_{DM}(k,z)=1$, is crossed for the scale 100 $h^{-1}$Mpc at redshift 0.24.}
\label{kselection:nz:density}
\end{center}
\end{figure}

In Fig. \ref{kselection:nz:density}, we compare the comoving density (in $h^3$Mpc$^{-3}$) for the entire $k<14$ galaxy sample, with the dark matter power spectrum. The linear component computed using CAMB \citep{Lewis:1999bs} and the non linear component measured in the Multidark simulations. We assume a constant bias of 1.5 as found in \citet{2015MNRAS.449..835R}.  The bias is slightly higher than the usual bias found for 2MASS galaxies due to the low-redshift cut applied, $z>0.07$, to construct the clustering sample \citep{2010MNRAS.406....2F}. The 2MASS $k<14$ galaxy sample has $n\, b^2\, P=1$ at $r=100\; h^{-1}$ Mpc for $z=0.24$. This galaxy sample is therefore useful to study the large-scale structure up to redshift $\sim0.2$; beyond that the density is too small to overcome shot noise. Note that the volume of the sphere with a radius $z<0.2$ is 0.78 Gpc$^3 h^{-3}$ and that the volume covered by our sample footprint with $|g_{lat}|>10$ is 0.64 Gpc$^3 h^{-3}$.

%%%%%%%%%%%%%%%%%%%%%%%%%%%%%%%%%
%			\section{Mock catalogs}
%%%%%%%%%%%%%%%%%%%%%%%%%%%%%%%%%
\section{MultiDark mock catalogs}
\label{sec:mock:catalog}

The BigMultiDark Planck simulation (BigMD, \citealt{2014arXiv1411.4001K}; \footnote{\url{www.multidark.org}}) was run adopting a Planck $\Lambda$CDM cosmology \citep{2013arXiv1303.5076P} using {\sc gadget-2} \citep{2005MNRAS.364.1105S} with 2.5 Gpc $h^{-1}$ on the side and $3840^3$ particles. Halos were identified based on density peaks including substructures using the Bound Density Maximum (BDM) halo finder \citep{1997astro.ph.12217K}.
We use BigMD to build the mock catalogs that will mimic the density of $k<14$ galaxies drawn from the 2MASS survey and described above. Given the strong gradient in galaxy density as a function of redshift (see Fig. \ref{kselection:nz}), we construct light-cones rather than use a single snapshot at the mean redshift of the sample; see below.

\subsection{k-selected sample corresponding halo population}
With our current BigMD mass resolution, it is not possible to mock the high galaxy number densities $>0.007\, h^3$Mpc$^{-3}$ observed at redshift $z<0.07$ (the resolution is too low). We therefore construct a mock catalog corresponding to a K-selected galaxy sample covering the redshift range $0.07<z<0.2$. Observationally, the total density of $k<14$ galaxies covering the redshift range $0.07<z<0.2$ is 14.2 deg$^{-2}$ or $0.0065\,h^3$Mpc$^{-3}$. The volume lost by excluding $z<0.07$ is 0.03 Gpc$^3 h^{-3}$ (i.e., 4.7\% of the total volume), leaving us with 0.61 Gpc$^3 h^{-3}$. Note that the line-of-sight comoving distance from redshift 0.07 is $d_C(z=0.07)=0.193 h^{-1}$ Gpc and from redshift 0.2 is $d_C(z=0.2)=0.572 h^{-1}$ Gpc.

We use 20 BigMD snapshots in the redshift range $0.07<z<0.2$, and divide each simulation in 8 sub boxes of 1.25 Gpc$/h$ on a side. In each of the 8 sub boxes we construct a light-cone using all 20 snapshots. 
We follow the standard procedure to construct light-cones from $N$-body simulations \citep{2005MNRAS.360..159B,2007MNRAS.376....2K}:
\begin{enumerate}
\item Set the properties of the light-cone: angular mask, radial selection function (number density) and number of snapshots within the redshift range covered.
We use each simulation snapshot within the redshift range to construct a slice of the light-cone; the redshift range of one slice from a snapshot at $z_i$ will be $((z_i+z_{i-1})/2,(z_i+z_{i+1})/2$.
\item We position an observer (at redshift $z=0$) in a point inside the box and we transform the coordinates of the box so that the observer is located at the Euclidean coordinates (x,y,z)=(0,0,0).
\item Sort all objects in each snapshot using the (sub)halo maximum circular velocity $V_{max}$. This step is necessary because more massive objects (brighter galaxies) will have a larger priority in our selection when we apply a cut to fix the number density in the slice (radial selection function). 
\item Transform Cartesian coordinates to RA, DEC and $r_c$ (comoving distance in real space), which we use to compute the redshift of each object using the equation
\begin{equation}
r_c(z)=\int\limits_0^z\frac{c dz'}{H_0\sqrt{\Omega_m(1+z')^3+\Omega_\Lambda}}.
\end{equation}
\item Select in each simulation snapshot (sub)halos such that $(z_i+z_{i-1})/2 < z < (z_i+z_{i+1})/2$ and $\bar{n}(v>v_{max})=\bar{n}(k<14,z)$ (Halo Abundance Matching, HAM). For that, we read the sorted objects until the expected number density is obtained. The angular selection function is a galactic longitude greater than 10: $|g_{lat}|>10$.
\item Compute the peculiar velocity along the line-of-sight for each object to compute its distance in redshift-space $\mathbf{s} =\mathbf{r}_c + (\mathbf{v}\cdot \hat{\mathbf{r}}) \hat{\mathbf{r}}/(aH)$.
\end{enumerate}
In this procedure, we assume that a $k$-selected sample is a stellar mass-selected sample and perform a halo abundance matching to reproduce the observed density of galaxies. This assumption is inaccurate in the sense that the true sample is not exactly complete in stellar mass. All the same, under this assumption, we derive the best possible BAO measurement in the local universe {\it i.e.} we provide a lower limit to the precision one can reach with the most stellar mass complete spectroscopic survey of the local universe.

\subsection{Comparison to data, small scale clustering}
Using the SDSS DR10 galaxy catalog \citep{2014ApJS..211...17A}, we can extract a galaxy sample that is representative of our $k<14$ magnitude limited sample by making the following cuts:
\begin{itemize}
\item $0.07<z<0.2$ and ZWARNING$\leq4$
\item $k<14$ (2MASS magnitude)
\item $130^\circ<ra<230^\circ$ and $10^\circ<dec<55^\circ$
\end{itemize}
This sample contains 55,022 galaxies with a mean redshift at 0.12, and covers 7396 deg$^2$ or a volume of $0.139$ Gpc$^3 h^{-3}$ with a density of 7 per deg$^2$. This average density is sufficient to overcome shot noise and measure the correlation function on scales smaller than 20 Mpc$h^{-1}$.% \simeq 70$h^{-1}$Mpc \times 70 $h^{-1}$Mpc \times 70 h^{-1}$Mpc if this were a box and therefore do not expect a truthful measurement of the autocorrelation beyond 20 Mpc$/h$. 

\begin{figure}
\begin{center}
\includegraphics[width=\columnwidth]{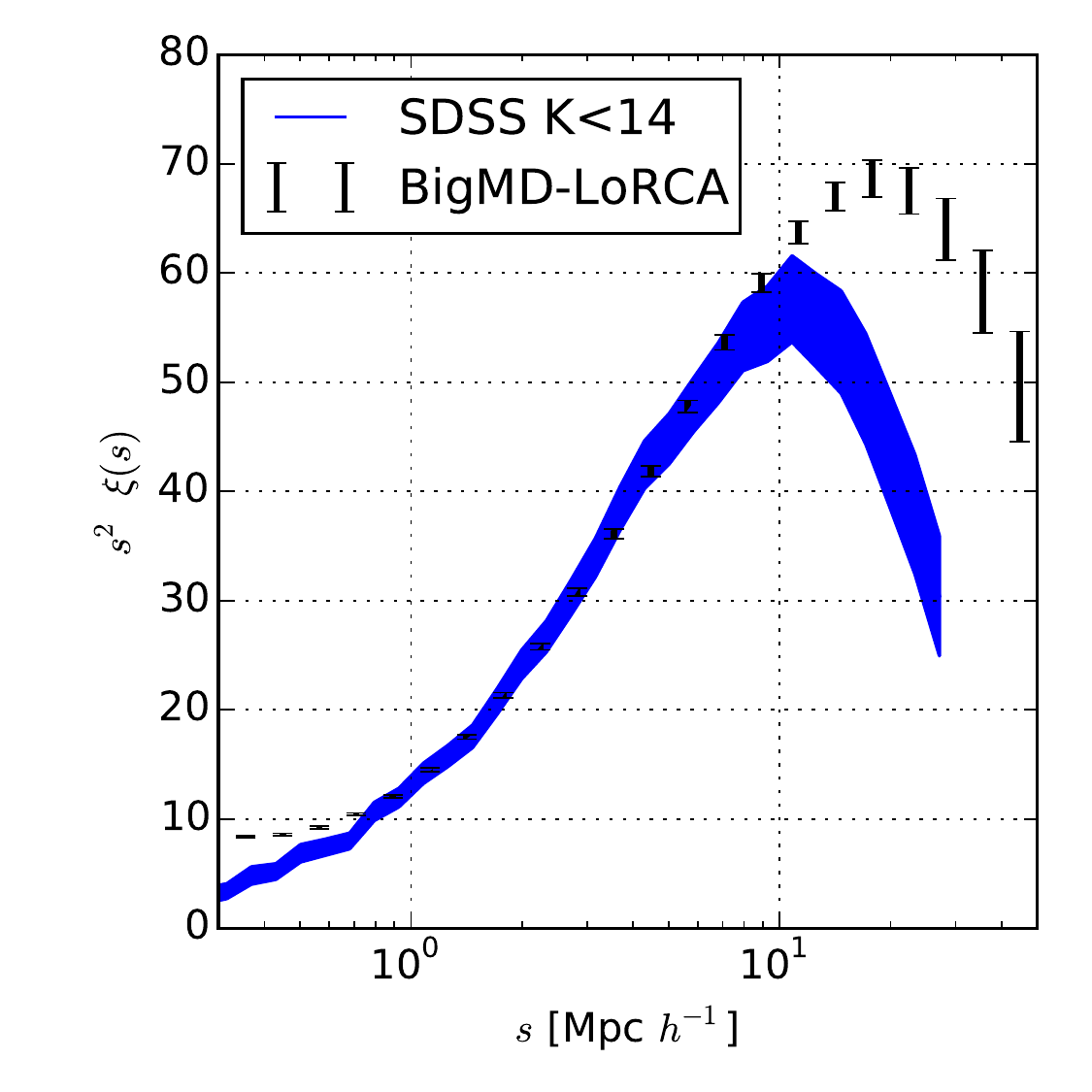}
\caption{Two-point correlation function of the SDSS $k<14$ galaxy sample compared to the mean and 1$\sigma$ errorbars of the BigMD mock light cones.}
\label{mock:data:comparison}
\end{center}
\end{figure}

We draw randomly from the SDSS mask which removes the bright stars and the unphotometric areas. 
%I assigned redshifts to the randoms by shuffling the true redshifts (no fits of the redshift distribution).
%I do not apply any weight to the data, as it looks like a fair sample of the complete k<14 mag limited sample. I do not convert it into a volume limited sample by selecting in absolute magnitude MK. I assume the light cone corresponds exactly to these galaxies.
In Fig. \ref{mock:data:comparison} we compare the clustering measurement to the mean of the 8 BigMultiDark mock light-cones. %Each of the 8 mocks contain 560 000 galaxies
The agreement between data and simulation is very good in the range $0.8 h^{-1}$Mpc - $10 h^{-1}$Mpc.
%F.P. There is only half error bars for the measurements at large distances ... 

We note two effects on smaller and larger scales. On smaller scales, we see the fiber collision effect, and hence no clustering enhancement on small scales below $s<0.8h^{-1}$Mpc. On larger scales, the clustering amplitude decreases at $s>20h^{-1}$Mpc due to the small volume of the sample, though the geometry allows for some large modes to be measured. 
Finally, this means the halo abundance matching models correctly reproduce the clustering of the galaxies we consider.

\subsection{Massive mock production}
\label{sec:ezmock}
We generate an additional 1,000 mock light cones using {\sc EZmocks} \citep{2015MNRAS.446.2621C} to construct the covariance matrix for the measurement of the two-point correlation function. {\sc EZmocks} have been designed to reproduce accurately, at very low computational cost, the large-scale structure of halos in $N$-body simulations in terms of the 2-point correlation function, power spectrum, 3-point correlation function, and bispectrum, in both real and redshift-space \citep[e.g. see][]{2015MNRAS.446.2621C}. Thus, {\sc EZmocks} is ideal for generating a large number of mock catalogues to construct a reliable covariance matrix for our local universe clustering measurements. To optimize computational resources, we derive the minimum number of snapshots necessary to make a truthful light-cone. We find that 6 redshift snapshots are sufficient to map the range $0.07<z<0.2$. Hence, we produce a 1,000 realizations of the six snapshots to construct each {\sc EZmock} light-cone of our local universe survey.

%%%%%%%%%%%%%%%%%%%%%%%%%%%%%%%%%
%			\section{BAO measruements}
%%%%%%%%%%%%%%%%%%%%%%%%%%%%%%%%%
\section{Cosmology with the BAO measurements from the BigMultiDark and EZmock light-cones}
\label{sec:BAOmeasurement}
In this section, we describe the measurements of the correlation function obtained
from the set of mock catalogues constructed using BigMD and {\sc EZmock}. We also show the results of the likelihood analysis that leads to a constraint on the measurement of the BAO peak. Since we intend to provide an estimation of the expected constraint on the BAO measurement from our galaxy sample, we ignore some factors, which might introduce small bias to the mean value of the correlation function but should have negligible effect on the uncertainty.

\subsection{Measuring the two-point correlation function}
\begin{figure}
\begin{center}
\includegraphics[width=\columnwidth]{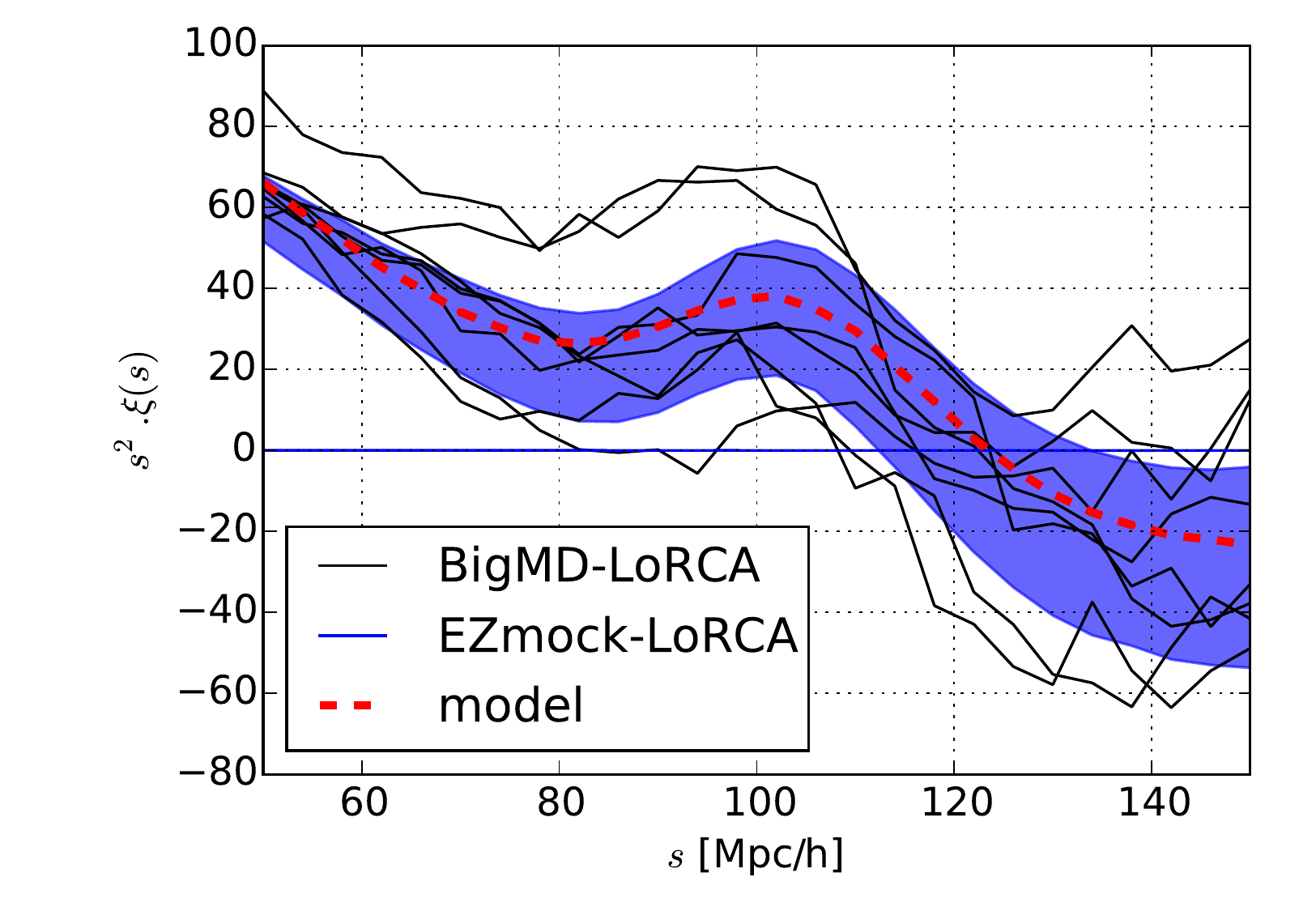}
\caption{The two-point correlation functions from the 8 BigMD-LoRCA mock catalogues (solid black lines) are compared with the $1\sigma$ region obtained from the 1,000 {\sc EZmocks} (blue region). The theoretical model is also shown (solid red).}\label{mocks:xis}
\end{center}
\end{figure}

We use the two-point correlation function estimator given by \citet{1993ApJ...412...64L},
%\begin{equation}
%\xi(s) = \frac{DD(s)-2DR(s)+RR(s)}{RR(s)},
%\end{equation}
%where DD, DR, and RR represent the normalized data-data,
%data-random, and random-random pair counts respectively in a
with a bin size of $4h^{-1}$Mpc. This estimator has minimal variance for a Poisson
process. The number of random points that we use is ten times that 
of the number of data points. We assign to 
each data point a radial weight of $1/[1+n(z)\cdot P_w]$ \citep{1994ApJ...426...23F}, where $n(z)$ is the radial selection function and $P_w = 2\cdot 10^4$ 
$h^{-3}$Mpc$^3$, as in \citet{2012MNRAS.427.3435A}.
Fig. \ref{mocks:xis} shows each of the correlation functions from the individual 8 BigMD mock catalogues, comparing with the mean correlation functions and 1-sigma error bar from the 1,000 {\sc EZmocks} described in Sec. \ref{sec:ezmock}.

\subsection{Measuring the covariance matrix}
\label{sec:covariance}
\begin{figure}
\begin{center}
\includegraphics[width=\columnwidth]{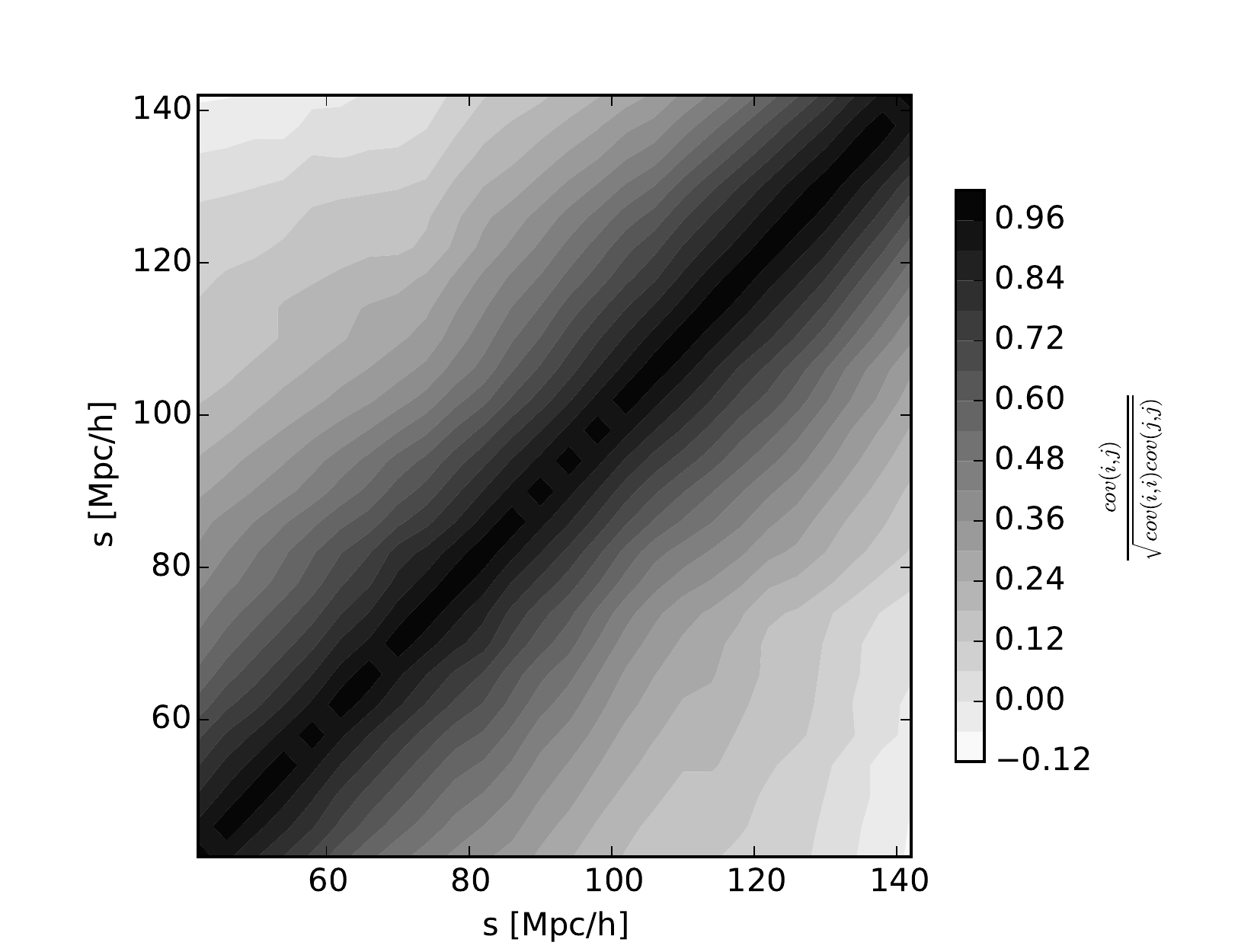}
\caption{Normalized covariance matrix based on the 1,000 {\sc EZmocks} coded after $O_{ij}=\frac{C_{ij}}{\sqrt{C_{ii}C_{jj}}}$ where $C_{ij}=\frac{1}{N-1}\sum^N_{k=1}(\bar{\xi}_i-\xi_i^k)(\bar{\xi}_j-\xi_j^k)$.}
\label{mocks:covar}
\end{center}
\end{figure}

We use the 1,000 {\sc EZmocks} light-cones described in Sec. \ref{sec:ezmock}
to estimate the covariance matrix of the correlation functions measured from the BigMD mock light-cones. We calculate the correlation functions
of the {\sc EZmocks} and construct the covariance matrix as follows

\begin{equation}
 C_{ij}=\frac{1}{N-1}\sum^N_{k=1}(\bar{\xi}_i-\xi_i^k)(\bar{\xi}_j-\xi_j^k),
\label{eq:covmat}
\end{equation}
where $N$ is the number of {\sc EZmock} catalogues, $\bar{\xi}_m$ is the
mean of the $m^{th}$ bin of the {\sc EZmock} correlation functions, and
$\xi_m^k$ is the value of the $m^{th}$ bin of the $k^{th}$ {\sc EZmock}
correlation function. Fig. \ref{mocks:covar} shows the normalized covariance matrix, $O$, which is defined by
\begin{equation}
 O_{ij}=\frac{C_{ij}}{\sqrt{C_{ii}C_{jj}}}.
\label{eq:covmat}
\end{equation}
%The covariance matrix can be described in mathematical terms as a one-to-one function with a Gaussian dispersion of 1$\sigma=21$ Mpc$/h$.
%

EZmock lightcone mocks have been calibrated with the simulation including the BAO signal. Therefore, they should be used for estimating the covariance matrix but not predicting the BAO significance. To obtain better forecast, one should use larger simulations or semi-N-body simulations (e.g. COLA, \citealt{2013JCAP...06..036T}) which are much more expensive than EZmocks; see \citet{2015MNRAS.446.2621C}

\subsection{Modeling the two-point correlation function}
To include the damping effects of nonlinear structure
formation and peculiar velocities at BAO scales, we calculate the
`dewiggled' power spectrum as follow,

\begin{equation}
P_{dw}(k)=P_{lin}(k)e^{\left(-\frac{k^2}{2k_\star^2}\right)}+P_{nw}(k)\left[1-e^{\left(-\frac{k^2}{2k_\star^2}\right)}\right],
\end{equation}
where $P_{lin}(k)$ is the linear power spectrum computed using CAMB
\citep{Lewis:1999bs}, $P_{nw}(k)$ is the no-wiggle power spectrum calculated using Eq.(29) in \cite{1998ApJ...496..605E}; we fix $k_\star=0.1$ in this analysis. 
In this study, we want to give an estimation of the expected constraint on the BAO measurement from our galaxy sample described above. 
Therefore, we do not use a more accurate model (e.g., see \citealt{2011MNRAS.417.1913R,2014MNRAS.437..588W,2014MNRAS.439.3630W}) 
which is more expensive for computation. 
These models take into account the nonlinear effects occurring at small scales to describe the full-shape of the correlation function.
Since we focus only on very large scales, i.e. $48 < s < 160h^{-1}$Mpc, to extract the BAO signal constraints from the monopole of the correlation function, the additional accuracy that may be provided by a more complicated model is small.

We recover the dewiggled two-point correlation function, $\xi_{dw}(s)$, by Fourier transforming the dewiggled power spectrum. Finally, the theoretical model of the correlation function is given by
\begin{equation} \label{eq:alpha}
\xi_{th}(s)=\xi_{dw}(\alpha s),
\end{equation}
where the rescaling parameter, $\alpha$, is referred to as the BAO measurement, since the major contribution of the constraint on $\alpha$ comes from the BAO peak signal.

\subsection{Likelihood Analysis}
The likelihood is taken to be proportional to $\exp(-\chi^2/2)$, and $\chi^2$ is given by
\begin{equation} \label{eq:chi2}
 \chi^2_{\rm mock}\equiv\sum_{i,j=1}^{N_{\rm bins}}\left[\xi_{\rm th}(s_i)-\xi_{\rm mock}(s_i)\right]
 C_{ij}^{-1}
 \left[\xi_{\rm th}(s_j)-\xi_{\rm mock}(s_j)\right],
\end{equation}
where $N_{\rm bins}$ is the number of bins used, $\xi_{\rm th}$ is the theoretical
 correlation function of our model, and $\xi_{\rm mock}$ is the correlation function measured from a BigMD mock light-cone. We use the scale range, $48 < s < 160h^{-1}$Mpc, with the bin size of $4h^{-1}$Mpc, so that $N_{\rm bins}=28$.

We use CosmoMC \citep{2002PhRvD..66j3511L} in a Markov Chain Monte-Carlo
likelihood analysis. The only parameter we explore is $\alpha$.
We marginalize over the amplitude of the correlation function and fix $\Omega_mh^2$, $\Omega_bh^2$, and $n_s$ to the input Planck cosmological parameters of the BigMD simulation.
 
\subsection{Results of BAO and broad-band shape measurements}
\label{sec:results}
Table \ref{table:BAO:fits} shows the individual $\alpha$ measurements and $\chi^2$ per degree of freedom (d.o.f.) obtained from each of the 8 full-sky BigMD mock light-cones. We also measure $\alpha$ for the average of the 8 mocks ($0.985$), and estimate the average of the their errors ($0.038\pm0.017$), and $\chi^2$/d.o.f. ($0.99$). One can see that the average $\chi^2$/d.o.f is very close to 1, which confirms that our covariance matrix is reasonable.The expected uncertainty on the BAO measurement for the proposed local universe sample is less than 4\%. See also in Table \ref{table:BAO:fits} the results for replacing the mock correlation function by the linear correlation function. Since there is no BAO damping in the linear correlation function, we then obtained an estimation for the expected precision on $\alpha$ provided a perfect BAO reconstruction (the best measurement one can reach). We obtain 1.2\%.

Therefore, even with this small volume survey, we conclude that a full-sky survey of galaxies with $k<14$ can reach a precision of 1.2\% - 4\% on the BAO standard ruler measurement. 

\begin{table}
\caption{The mean and standard deviation of the $\alpha$ parameter for each of the 8 individual full-sky BigMD light-cones. We also show the average of the means and the average of the standard deviations. The last line shows the fit obtained by replacing the mock correlation function with the linear correlation function.}
\begin{center}
\begin{tabular}{l c c c }
\hline \hline
Full sky & $\alpha$ & $\sigma_\alpha$&$\chi^2/$d.o.f \\
\hline
Mock 1 & 1.001 & 0.051 &0.60 \\
Mock 2 & 1.013 & 0.035 &0.90 \\
Mock 3 & 0.939 & 0.029 &1.06 \\
Mock 4 & 0.979 & 0.043 &0.93 \\
Mock 5 & 0.966 & 0.022 &1.13 \\
Mock 6 & 1.059 & 0.063 &0.80 \\
Mock 7 & 0.936 & 0.027 &1.49 \\
Mock 8 & 0.986 & 0.032 &1.01 \\ \hline
Average& 0.985 & $0.038\pm0.016$ &0.99 \\ 
Linear CF & 1.001 &  0.012 & \\
\hline
\end{tabular}
\end{center}
\label{table:BAO:fits}
\end{table}%

Now, we cut each of the mocks in two or four to estimate the expected determination of $\alpha$ when using only half-sky (6dF survey-like) or a quarter-sky (SDSS survey-like). The results are summarized in Tables \ref{table:6df} and \ref{table:SDSS}. With half-sky (quarter-sky) survey, the results show that one can measure $\alpha$ at the 7 percent precision level (12 percent), which could be reduced by using a perfect reconstruction technique to 2.4 percent level (5.3 percent). \citet{2011MNRAS.416.3017B} measured BAO (without using reconstruction) with 4.5\% precision using the 6dF survey, and \citet{2015MNRAS.449..835R} measured BAO using the main SDSS survey at the 3.8\% precision level using reconstruction. Considering our analysis, both error estimates seems to be too optimistic.

\begin{table} 
 \caption{Same as Table \ref{table:BAO:fits}, but uses 6dF survey-like mock light-cones for half-sky.}
 \begin{center}
 \begin{tabular}{l c c c} 
     \hline
Half-sky & $\alpha$ & $\sigma_\alpha$&$\chi^2/$d.o.f \\
     \hline
     6dF mock $1$ & $0.959$ & $0.071$ & $0.89$ \\ 
     6dF mock $2$ & $0.957$ & $0.083$ & $1.02$ \\
     6dF mock $3$ & $0.948$ & $0.064$ & $0.72$ \\
     6dF mock $4$ & $1.038$ & $0.055$ & $1.65$ \\
     6dF mock $5$ & $0.948$ & $0.046$ & $0.75$ \\
     6dF mock $6$ & $0.927$ & $0.083$ & $1.14$ \\
     6dF mock $7$ & $1.102$ & $0.078$ & $0.77$ \\
     6dF mock $8$ & $0.987$ & $0.077$ & $0.68$ \\
     6dF mock $9$ & $0.958$ & $0.033$ & $0.69$ \\
     6dF mock $10$ & $0.939$ & $0.057$ & $0.79$ \\
     6dF mock $11$ & $0.992$ & $0.058$ & $0.60$ \\
     6dF mock $12$ & $1.107$ & $0.062$ & $1.58$ \\
     6dF mock $13$ & $0.922$ & $0.056$ & $1.13$ \\
     6dF mock $14$ & $0.951$ & $0.101$ & $0.61$ \\
     6dF mock $15$ & $1.063$ & $0.079$ & $0.98$ \\
     6dF mock $16$ & $0.959$ & $0.083$ & $0.53$ \\
    \hline 
     Average & $0.985$ & $0.068\pm0.017$ & $0.91$ \\
     Linear CF & $1.002$ &  $0.024$ &  \\
     \hline
 \end{tabular}
 \end{center}
\label{table:6df}
  \end{table}

 \begin{table}
\caption{Same as Table \ref{table:BAO:fits}, but uses SDSS survey-like mock light-cones for a quarter-sky.}
 \begin{center}
  \begin{tabular}{l c c c} 
     \hline
Quarter-sky & $\alpha$ & $\sigma_\alpha$&$\chi^2/$d.o.f \\ 
     \hline
    SDSS mock $1$ & $1.019$ & $ 0.075$ & $0.87$ \\ 
    SDSS mock $2$ & $1.01$ & $0.16$ & $1.76$ \\
    SDSS mock $3$ & $1.06$ & $0.13$ & $0.70$ \\
    SDSS mock $4$ & $1.06$ & $0.14$ & $0.81$ \\
    SDSS mock $5$ & $1.14$ & $0.11$ & $0.53$ \\
    SDSS mock $6$ & $0.97$ & $0.14$ & $0.87$ \\
    SDSS mock $7$ & $1.048$ & $0.069$ & $1.38$ \\
    SDSS mock $8$ & $1.05$ & $0.10$ & $0.68$ \\
    SDSS mock $9$ & $1.028$ & $0.087$ & $1.10$ \\
    SDSS mock $10$ & $1.03$ & $0.12$ & $0.79$ \\
    SDSS mock $11$ & $0.92$ & $0.12$ & $1.02$ \\
    SDSS mock $12$ & $1.02$ & $0.19$ & $0.81$ \\
    SDSS mock $13$ & $0.99$ & $0.16$ & $0.94$ \\
    SDSS mock $14$ & $1.11$ & $0.18$ & $0.80$ \\
    SDSS mock $15$ & $0.99$ & $0.13$ & $0.76$ \\
    SDSS mock $16$ & $1.01$ & $0.12$ & $0.76$ \\
    SDSS mock $17$ & $1.016$ & $0.071$ & $1.03$ \\
    SDSS mock $18$ & $1.01$ & $0.15$ & $1.33$ \\
    SDSS mock $19$ & $1.02$ & $0.11$ & $0.79$ \\
    SDSS mock $20$ & $1.01$ & $0.17$ & $1.03$ \\
    SDSS mock $21$ & $1.064$ & $0.072$ & $0.93$ \\
    SDSS mock $22$ & $1.013$ & $0.065$ & $1.52$ \\
    SDSS mock $23$ & $1.16$ & $0.13$ & $1.31$ \\
    SDSS mock $24$ & $1.06$ & $0.15$ & $1.81$ \\
    SDSS mock $25$ & $0.912$ & $0.086$ & $1.61$ \\
    SDSS mock $26$ & $0.929$ & $0.074$ & $1.30$ \\
    SDSS mock $27$ & $1.05$ & $0.12$ & $1.27$ \\
    SDSS mock $28$ & $1.00$ & $0.14$ & $1.07$ \\
    SDSS mock $29$ & $1.01$ & $0.13$ & $0.63$ \\
    SDSS mock $30$ & $1.14$ & $0.12$ & $0.99$ \\
    SDSS mock $31$ & $0.96$ & $0.15$ & $0.81$ \\
    SDSS mock $32$ & $1.01$ & $0.11$ & $1.02$ \\
    \hline
    Average & $1.026$ & $0.12\pm0.03$ & $1.03$ \\
    Linear CF & $1.014$ & $0.053$ &  \\
     \hline
 \end{tabular}
 \end{center}
 \label{table:SDSS}
  \end{table}

\subsection{Broad-band and systematics effects}
To study the impact of systematics on the BAO fits, we introduce a broad band term in the model of the 2PCF to mimic the effects of systematics :
\begin{equation}
\rm \xi(r, with\; systematics)=a_0+\frac{a_1}{r}+\frac{a_2}{r^2}+\xi(r,no\; systematics)
\end{equation}
Using the full-sky mocks, we perform the likelihood analysis with two sets of priors: $-0.0015<a_0< 0.0015$, $-1.5     <a_1<  1.5$ and $-10     <a_2<   10$ and the double (second set). In this way, one can check whether the marginalizing over the systematic priors have an effect on the final measurement of alpha \citep{2014MNRAS.445....2V}. We find that the precision on alpha is slightly degraded from 3.9\% (no systematics) to 4.1\% (first set of priors) and 4.9\% (second set of priors); see Table \ref{result:with:syst}.

 \begin{table*}
 \caption{\label{result:with:syst} The mean and error on the $\alpha$ parameter for each of the 8 individual full-sky BigMD light-cones including systematics.}
 \begin{center}
  \begin{tabular}{c c c c c c c} 
     \hline
     $  $ & $\alpha$(no systematics)& $\chi^2/d.o.f.$ & $\alpha$(with systematics priors) & $\chi^2/d.o.f.$ & $\alpha$ (with double priors) & $\chi^2/d.o.f.$ \\ \hline \hline 
 mock 1
  & $      1.001 \pm      0.051   $  
& $   0.60 $
  & $      0.997 \pm      0.062   $ 
  & $ 0.683   $
 & $      1.002 \pm      0.075   $ 
  & $ 0.685   $
 \\
 mock 2
  & $      1.013 \pm      0.035   $  
& $   0.90 $
 & $      1.011 \pm      0.039   $ 
  & $ 1.025   $
 & $      1.011 \pm      0.042   $
  & $ 1.017   $ 
 \\
 mock 3
  & $      0.939 \pm      0.029   $ 
& $   1.06 $
 & $      0.932 \pm      0.038   $ 
& $ 0.956   $
 & $      0.937 \pm      0.050   $
& $ 0.819   $ 
 \\
 mock 4
  & $      0.979 \pm      0.043   $ 
& $   0.93 $
 & $      0.967 \pm      0.041   $ 
& $ 1.014   $
 & $      0.967 \pm      0.053   $
& $ 1.017   $
 \\
 mock 5
  & $      0.966 \pm      0.022   $ 
& $   1.13 $
 & $      0.956 \pm      0.022   $ 
& $ 0.980   $
 & $      0.953 \pm      0.027   $
& $ 0.986   $
 \\
 mock 6
  & $      1.059 \pm      0.063   $ 
& $   0.80 $
 & $      1.055 \pm      0.063   $ 
& $ 0.747   $
 & $      1.049 \pm      0.063   $
& $ 0.693   $
 \\
 mock 7
  & $      0.936 \pm      0.027   $ 
& $  1.49  $
 & $      0.946 \pm      0.026   $ 
& $ 1.537   $ 
 & $      0.944 \pm      0.027   $
& $ 1.537   $
 \\
 mock 8
  & $      0.986 \pm      0.032   $  
& $  1.01  $
 & $      0.989 \pm      0.044   $ 
& $ 1.108   $ 
 & $      0.993 \pm      0.057   $
& $ 1.112   $
 \\
\hline
 Average 
 & $0.985\pm 0.038$
 & $  0.99  $
 & $0.982\pm 0.041$
 & $  1.01  $
 & $0.982\pm 0.049$
 & $  0.98  $
 \\
     \hline
 \end{tabular}
 \end{center}
  \end{table*}

\subsection{Impact of redshift errors}
\label{subsec:redshiftERR}
In the previous sections, we assumed a perfect measurement of the redshift.
We investigate the impact of eventual redshift errors on the estimation of the two-point correlation function. To this aim, we degrade for example the mock number 1 into observed mocks as follows:
\begin{itemize}
\item We simulate {\bf redshift errors} by drawing observed redshifts from a Gaussian distribution centered on the exact redshift with a scatter $dz$, i.e. $z_{obs}=\mathcal{N}(z_{true},dz)$ with different values of $dz$: 0.0001, 0.0005, 0.001, 0.005, 0.01. For each value of dz, we produce 50 observed mocks on which we measure the correlation function. We take the mean of the 50 correlation functions as that obtained including redshift errors.
In the scale range from 40 to 110 $h^{-1}$Mpc, the deviations from the original correlation function is smaller than 0.5\% for dz=0.00005, 1\% for dz=0.0001, 2\% for dz=0.0005, 4\% for dz=0.001, 25\% for dz=0.005, and within a factor of 3 for dz=0.01. %See Fig. \ref{observed:xis} left panel.
Hence, we conclude that for dz$\geq$0.005, the discrepancy between the observed and the true correlation function is larger than the statistical error predicted by the mocks.
A reasonable aim for the spectroscopic redshift precision of a local universe BAO survey is dz$\leq$0.001, which can be achieved by a $R\geq1000$ spectrograph.
\item We simulate {\bf catastrophic redshift errors} by shuffling a certain percentage of redshifts: 0.5, 1, and 5\%. On large scales, catastrophic redshifts have an impact below the percentage level if the fraction is below 1 percent. On small scales, the clustering is underestimated by about 2\% for 1\% fraction of catastrophic redshifts.% See Fig. \ref{observed:xis} middle panel.
The main effect is that such errors damp the clustering signal at all scales. The fluctuations of the correlation function measurement due to such catastrophic redshift errors remains within the statistical errors.
\item We simulate {\bf systematic redshift errors} by adding a fixed offset to all the redshifts, i.e. $z_{obs}=z_{true}+ {\it offset}$, which takes the values 0.00005, 0.0001, 0.0005, and 0.001. This systematic redshift error strongly affects the measurement of the two-point correlation function. Even the smallest systematic error has a two percent impact on large scales. The fluctuations of the correlation function estimation due to such systematic redshift errors remains within the statistical errors. Similarly to the catastrophic redshift errors, such uncertainty damps the signal at all scales.
\end{itemize}
In summary, the redshift errors described above would lead to a worse estimation of the two-point correlation function, and may introduce a damping of the BAO peak, but they do not shift the BAO scale.

The 2MASS full sky photometric redshift catalog \citep{2014ApJS..210....9B} currently has redshift errors of 0.015 which is much larger than the 0.005 required. It is therefore timely to obtain spectroscopic redshifts of the full sky to extract all the cosmological information (only 32\% were observed to date).

%%%%%%%%%%%%%%%%%%%%%%%%%%%%%%%%%%%%%%%%%

\subsection{Probing Dark Energy}% $\Lambda$CDM model}
\label{subsec:wDE}

\begin{figure}
\begin{center}
\includegraphics[width=1.0\columnwidth]{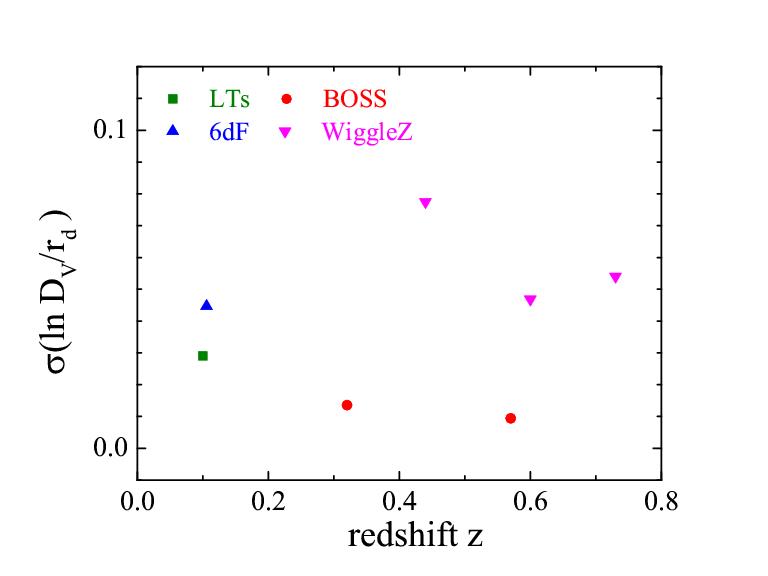}
\caption{Relative error the distance (\%) vs redshift. LTs will provide a 3.8\% measurement LoRCA+Taipan (green square). It is compared to 6dF (blue triangle, \citealt{2011MNRAS.416.3017B}), BOSS (red circle, \citealt{2015arXiv150906371C}) and WiggleZ (\citealt{Blake01092012}).}\label{fig:Distance:redshift}
\end{center}
\end{figure}

Probing cosmology is one of the key science drivers of most existing and upcoming redshift surveys, especially for the LoRCA + TAIPAN survey (LTs). We project the forecast on a Hubble diagram in the context of the $\Lambda$CDM paradigm; see Fig. \ref{fig:Distance:redshift}. The improvement compared to previous BAO, SNe is clear.

Aubourg et al. (2014) combined CMB, BAO and SNe, leaving curvature as a free parameter, and reported for a constant equation of state for DE $w=-0.98\pm0.06$. Allowing for a redshift varying equation of state, although the significance is not yet strong enough to be conclusive, it shows a hint of dynamical dark energy (DDE) \citep{wrecon}. In this section, we perform a forecast for the sensitivity of LTs to probe the dark energy, in comparison with that of the SDSS-III BOSS survey \citep{BOSSDR12}.

\subsubsection{Confirming the Fisher formalism}
To study how DDE can be constrained with LTs, we use the conventional Fisher matrix method \citep{Fisher}. We first forecast the sensitivity of the volume averaged distance $D_V(z)$ following \citet{FisherPk} for LTs and BOSS, where 
\begin{equation} 
  D_V(z)\equiv\left[\frac{cz(1+z)^2D_A^2(z)}{H(z)} \right]^{1/3} 
\end{equation}
Here $D_A(z)$ and $H(z)$ are the comoving angular diameter distance and the Hubble parameter at redshift $z$ \footnote{Note that the fractional uncertainty of $D_V$ is the same as that of $\alpha$, i.e., $\sigma_{D_V}/{D_V}=\sigma_\alpha/\alpha$}. %The result is shown in Fig \ref{fig:BAO}. 
The Fisher BAO forecast for LTs is consistent with the mock result, i.e., the sensitivity is $\sim4\%$ at the effective redshift. We hereby confirm the accuracy of Fisher matrix forecasts and use it to project the uncertainty of the BAO scale onto that of dark energy parameters \footnote{This is the result using BAO alone, with all other cosmological parameters fixed at their fiducial values.}. 

\subsubsection{Constraints on DDE}

We consider two different configurations for $w(z)$, the equation-of-state (EoS) of dark energy, \begin{eqnarray}
{\rm I.}~~~~w(a)&=&w_0+w_a(1-a) \\
{\rm II.}~~~~ w(a)&=&w_i~(a_i<a<a_{i+1})
\end{eqnarray}

Configuration I, the so-called CPL configuration \citep{L,CP}, is widely used in the literature for parameter constraints due to its transparent physical meaning ($w_0=w(a=1),~w_0+w_a=w(a=0),~w_a=-dw/da$) and simplicity. However, it cannot capture any dynamics of $w(z)$ beyond the linear order of $a$. Configuration II is more general as $w$ is essentially a freeform function. In this work, we use 40 bins, uniform in the scale factor from $a=1.0$ to $a=0.5$, and assume that $w$ is a piecewise constant within each bin. The advantage of this configuration is that it allows model-independent study of $w(z)$, such as the principle component analysis (PCA), which has been applied to dark energy studies \citep{wPCA1,wPCA2,wPCA3}. 

\begin{figure}
\begin{center}
\includegraphics[width=1.0\columnwidth]{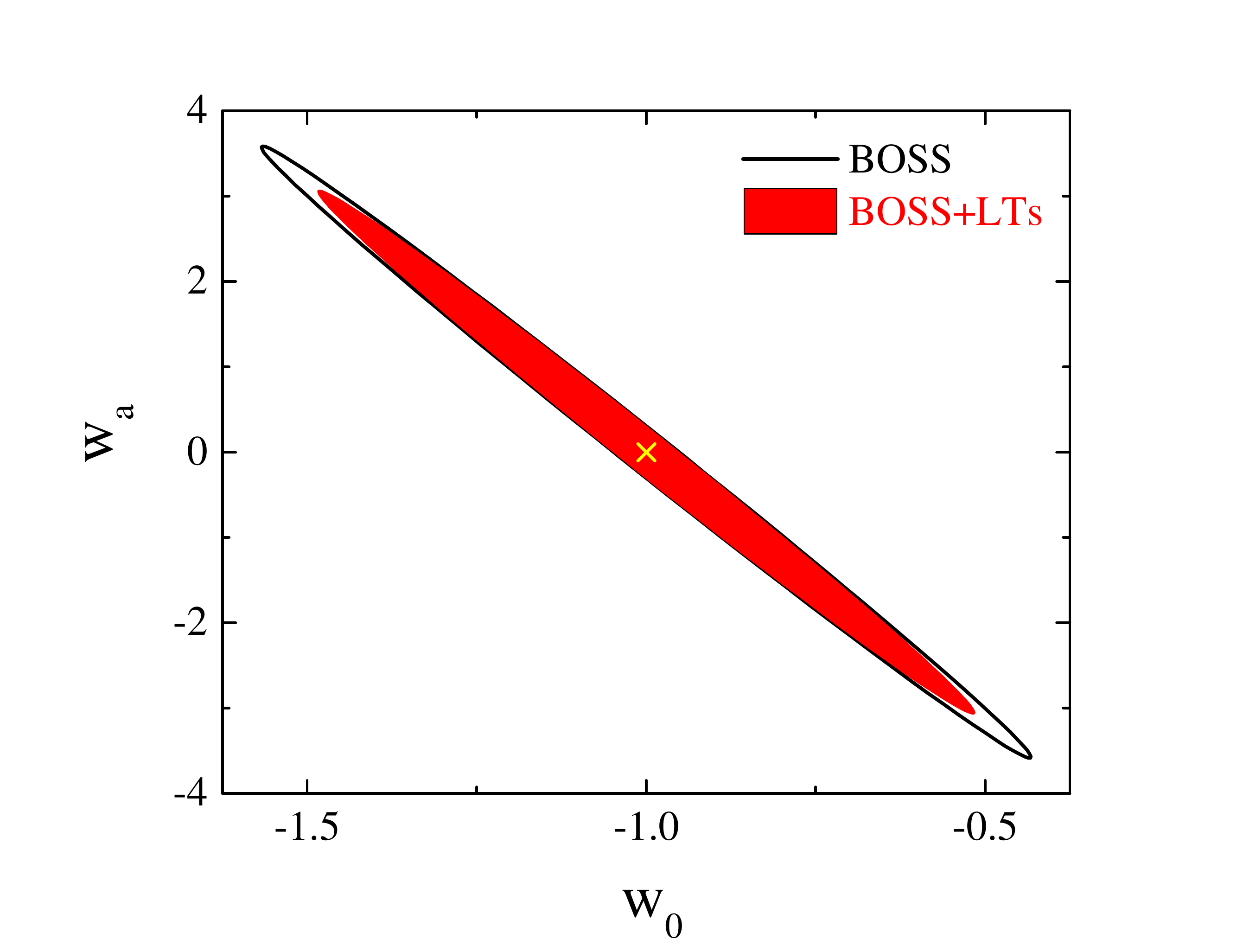}
\caption{The 68\% CL contour plot for $w_0,w_a$ using BOSS BAO data (black) and LTs (red filled) respectively. The cross denotes the fiducial model, which is the $\Lambda$CDM model.}
\label{fig:CPL}
\end{center}
\end{figure}

The result for the CPL configuration is shown in Fig~\ref{fig:CPL}. As seen, adding in LTs improves the constraints on $w_0$ and $w_a$ by 15\% and 14\% respectively. Note that although LTs has no high-$z$ ability, it can still improve the constraint on $w_a$ by breaking the parameter degeneracy. 
According to the Figure of Merit (FoM), as defined by the DETF \citep{DETF}, the improvement is 17\%.

Next, we move to the constraints on the binned $w(z)$, which allows more degrees of freedom. Although data cannot constrain any individual bin, some linear combinations of the bins can be well constrained. The eigenmodes, which define the mapping between the linear combinations and the original bins, can provide important information such as where the `sweet-spots' (the redshift at which the uncertainty of $w(z)$ gets minimized) are, and how many parameters can be well constrained, regardless of the form of configuration. The eigenmodes also provide a natural orthonormal basis onto which any $w(z)$ can be expanded, i.e., 
\begin{equation}
w(z)+1=\sum_i c_i e_i(z)
\label{eqn:eigenvec}
\end{equation} 
where $e_i(z)$ denotes the $i$th eigenvector and $c_i$ is the corresponding coefficient for the expansion. The eigenmodes can be found by diagonalising $F$, the Fisher matrix of the $w$ bins, \begin{equation}
F=W^T~\Lambda~W
\end{equation} The rows of $W$ store the eigenvectors and the diagonal matrix $\Lambda$ records the inverse error of the $c_i$'s. For more details of this PCA prescription, we refer the readers to \citet{wPCA1,wPCA2,wPCA3}. 

\begin{figure}
\begin{center}
\includegraphics[width=1.0\columnwidth]{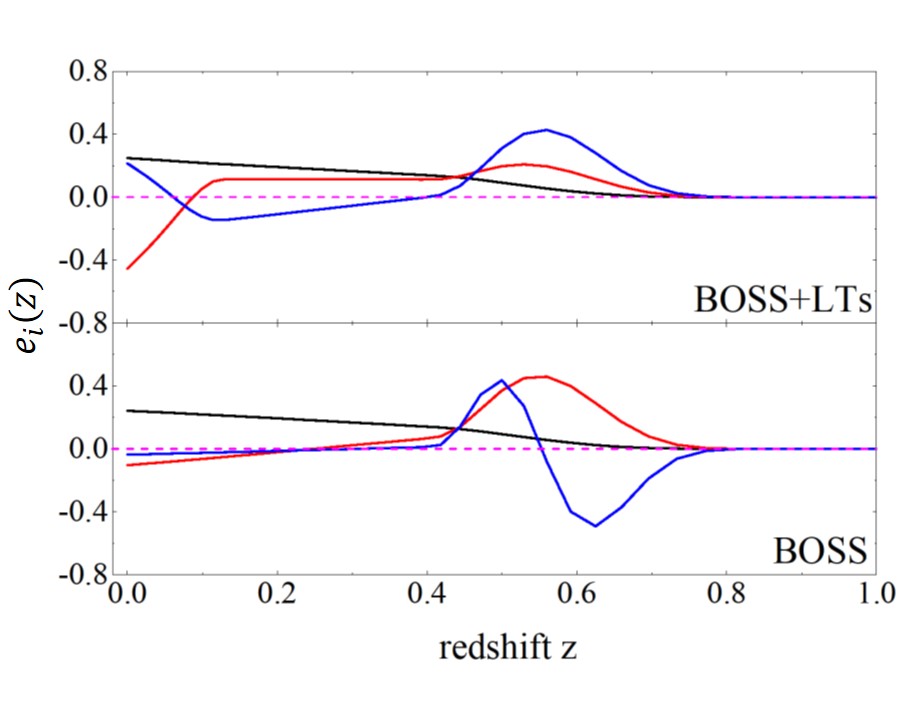}
\caption{The first three best constrained eigenvector, defined in equation (\ref{eqn:eigenvec}), using different combination of the data BOSS+LTs survey (upper panel) compared to BOSS alone (lower panel). The modes are shown, in the order from better constrained to worse as black, red and blue curves. The short dashed horizon line shows $e_i(z) = 0$.
4.3}
\label{fig:evec}
\end{center}
\end{figure}

\begin{figure}
\begin{center}
\includegraphics[width=0.8\columnwidth]{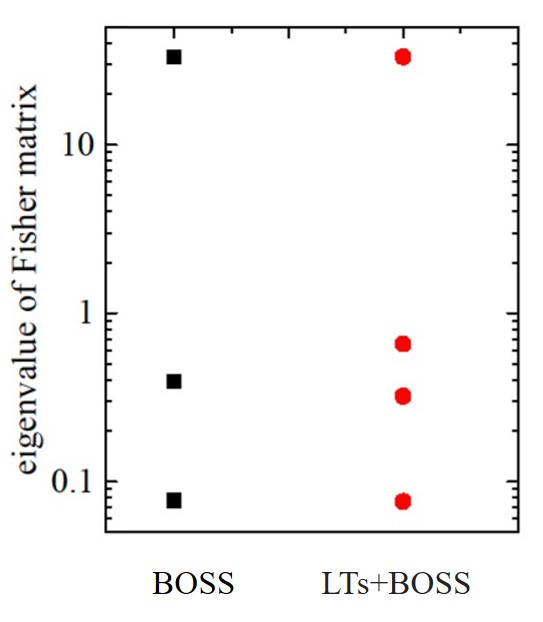}
\caption{The eigenvalues of the BAO Fisher matrix for BOSS (black squares) and BOSS+LTs survey (red circles). Adding LTs to BOSS, one constrains one more mode.}
\label{fig:eval}
\end{center}
\end{figure}

The PCA result is shown in Figs \ref{fig:evec} and \ref{fig:eval}. Fig \ref{fig:evec} shows the first three best constrained eigenmodes using BOSS and BOSS+LTs respectively. BOSS alone does not have any sensitivity against the variation of $w(z)$ at low redshift. However, when LTs is combined, the low-$z$ sensitivity (at $z<0.2$) is gained starting from the second eigenmode. The constraints on these modes are shown in \ref{fig:eval}. As shown, LTs improves the constraint on the second mode due to its low-$z$ reach. 

In summary, covering the low redshift ranges, LTs provides complementary dark energy constraints to BOSS. Assuming a CPL configuration of $w(z)$, LTs improves the FoM by 17\%. A more general PCA approach reveals that LTs helps detect the dynamics of dark energy at $z<0.2$, which is crucial to dark energy studies \citep{lowzDE}.

%%%%%%%%%%%%%%%%%%%%%%%%%%%%%%%%%%%%%%%%%
\section{LORCA: the low redshift survey at calar alto}
\label{sec:LORCA}
Considering this study, we propose the Low Redshift survey at Calar Alto (LoRCA) that aims observing $\sim 200,000$ galaxies to complete the sample described above of $\sim 800,000$ $k<14$ galaxy redshifts in the northern sky.

Such a survey would constitute the northern counterpart of the first year of the Australian TAIPAN survey.
Together, both surveys would reach the BAO precision limit reported 
in the previous section, and both LoRCA and TAIPAN will buildup the ultimate 
bright galaxy survey of the nearby universe. See next section for 
other interesting science cases along with the BAO measurements. 

This survey would use the Calar Alto 80-cm Schmidt 
telescope\footnote{\url{http://www.caha.es/CAHA/Telescopes/schmidt.html}} refurbished with a cartridge that hosts a plug aluminium plate that sits in the focal plane. %Optical fibers plugged into the holes in the plate carry light from the focal plane to the slit head, which feeds one spectrograph permanently mounted on the floor in a control environment room. A set of two plug-plates will be interchangeable. 

The field-of-view that we intend to cover with a single plate is 24 cm by 24 cm on the focal plane, i.e. 5.5$^\circ \times$5.5$^\circ$. We plan to use 400 fibers robots to acquire the light from the targets. Fiber cables will measure about 15-m long. The fiber core will be 100 microns, which yield 8.6 arc second on the telescope's focal plane. 

We will use an existing spectrograph with sufficient (see below) resolving power R=1,100 that covers the wavelengths $3,200 <\lambda< 7,300{\rm \AA}$, that allows us to pack all 300 fibers on the pseudo-slit and map them on a 2k by 2k detector. Such resolution would allow estimation of the redshifts with an error better than dz$\leq$0.001.
 
\begin{figure*}
\begin{center}
\includegraphics[width=0.95\textwidth]{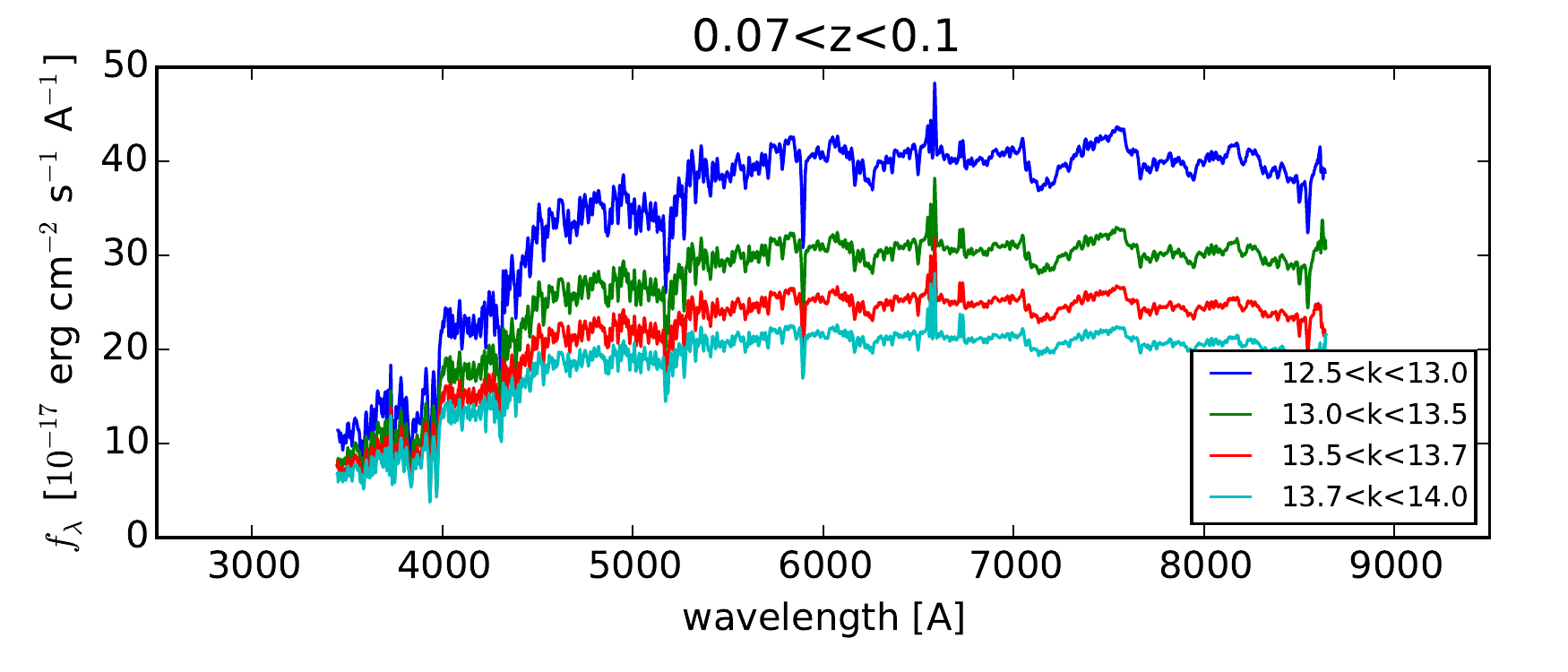}
\includegraphics[width=0.95\textwidth]{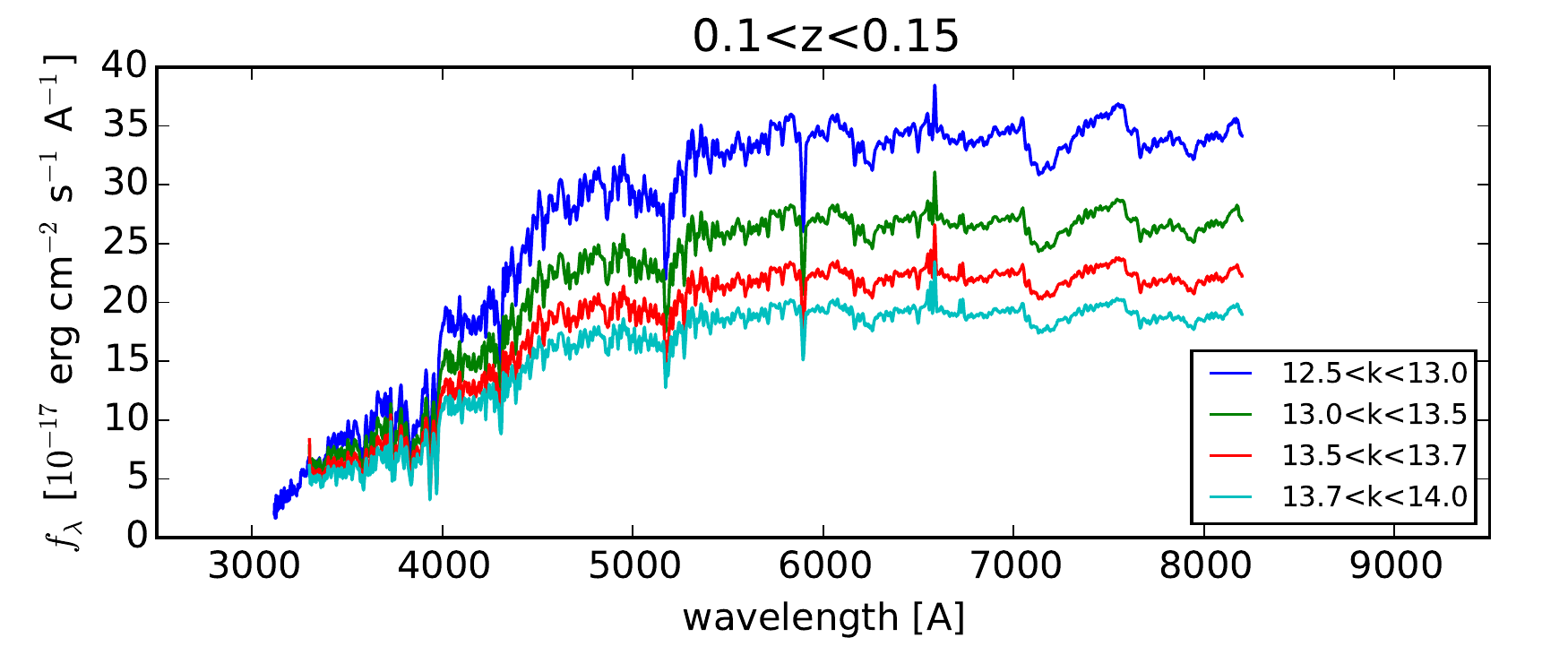}
\includegraphics[width=0.95\textwidth]{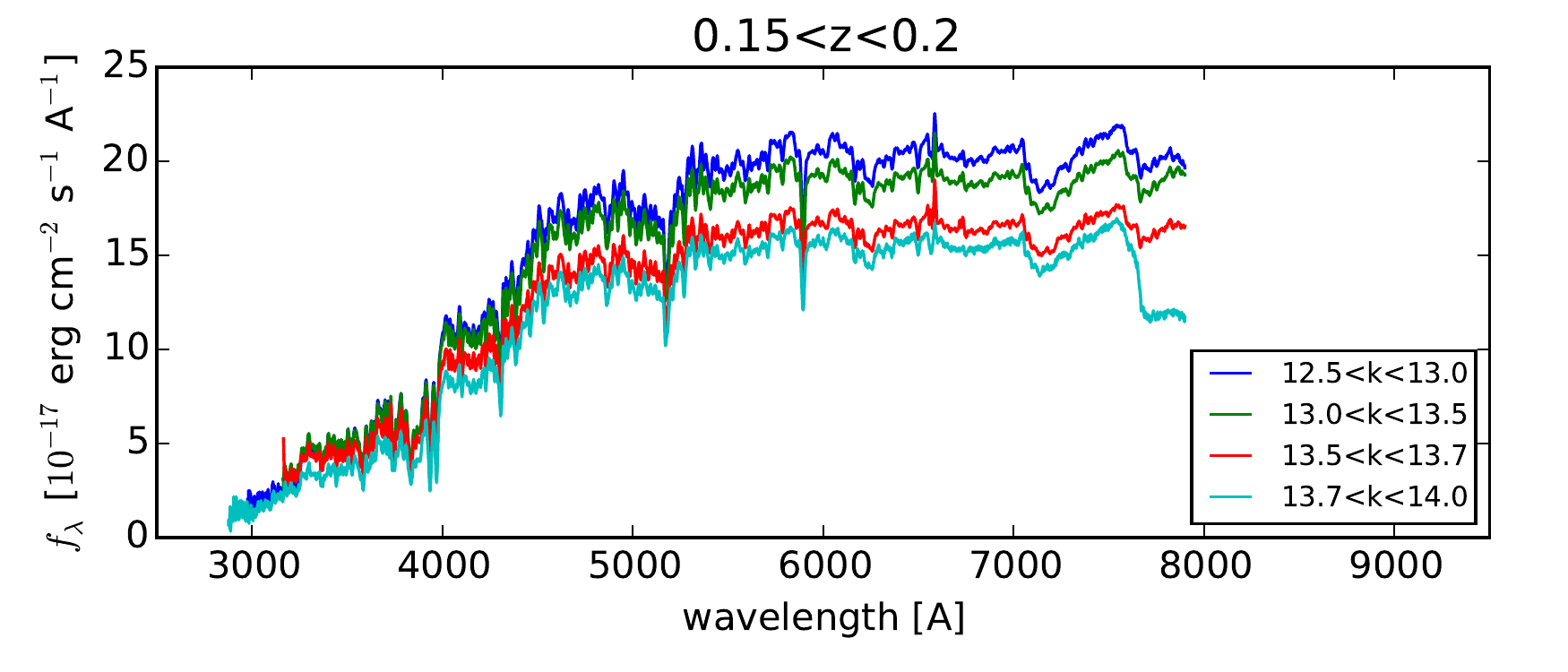}
\caption{Typical SDSS spectra binned in redshift and K band magnitude to be observed by the $k<14$ LoRCA bright galaxy survey.  In all panels, we show four mean stacked spectra corresponding to the magnitude bins: $12.5<K<13$, $13<K<13.5$, $13.5<K<13.7$, $13.7<K<14$ from bottom to top. Note the difference in the y-axis scale.}
\label{galaxy:stacks:BG}
\end{center}
\end{figure*}

 %\subsubsection*{Spectral library, redshift and exposure time}
Given that the SDSS data base \citep{2014ApJS..211...17A} already contains a 
fair spectroscopic sample of the galaxies to be observed, using a stacking procedure, 
we produced a library of high signal-to-noise ratio typical spectra of the LoRCA bright galaxy sample. We stacked the available SDSS spectra in three bins of redshifts and four bins of 2MASS K band magnitude; see Fig.\ref{galaxy:stacks:BG}. In the first redshift bin $0.07<z<0.1$ (top panel), the 4,000 \AA$ $ Balmer-break is deep and the H$\alpha$ emission line is well detected. In the highest redshift bin $0.15<z<0.2$ (bottom panel), 
the Balmer-break feature is about two times weaker, though still very well defined, but the H$\alpha$ line becomes weaker.  The fainter $13.7<k<14$ galaxies in the highest redshift bin should therefore drive the exposure time calculations. Note that the amount of features visible in absorption are sufficient to obtain reliable redshifts in $\sim3.5$ hour exposures (signal-to-noise ratio above 10). Thus, it is not necessary to have a spectrograph that samples out to the H$\alpha$ line. 
A sufficient wavelength coverage would be 3600$\,$\AA$\,$to 7200$\,$\AA, which is included in the currently available spectrograph (H$\alpha$ goes out of the spectrograph at redshift 0.09).

With the observational setting summarized in Table \ref{table:lorca:survey:param}, we can plan the observation of 2 plates per night, including overhead of field acquisition and necessary calibrations. We aim to complete the observation of 200,000 redshifts with 700 plates in a 3-year survey that will cover half of the northern sky, starting 2016 (1 year after the start of TAIPAN), observing in dark time only and taking into account the average time lost due to bad weather at Calar Alto.
\begin{table}
\caption{Summary of the parameters of LoRCA.}
\begin{center}
\begin{tabular}{c c}
\hline
Wavelength coverage & $3,600 <\lambda< 7,300$ \AA \\
Resolution & 1,100 \\
Fiber aperture & 8.6 arc seconds \\
Selection & $k_{2MASS}<14$ \\
N redshifts & 200,000 \\
N plates & 700 \\
\hline
\end{tabular}
\end{center}
\label{table:lorca:survey:param}
\end{table}%

\section{Additional cosmological probes with LoRCA}
With the observational parameters specified for the BAO study, LoRCA can also address other key questions in cosmology.

\label{sec:AddCosmoProbes}
\subsection{Peculiar velocities}
In the cosmology community there is a steadily increasing interest in measurements of peculiar velocities of galaxies. \citet{2014MNRAS.445.4267K} made a conclusive study discussing how new peculiar velocity surveys will be competitive as cosmological
probes.  Peculiar velocity can improve for example the growth rate constraints by a factor two (and up to five) compared to galaxy density alone.%

The immediate use of peculiar velocities is for cosmic flows mapping in the Local Volume \citep{2013AJ....146...69C}. Large coherent flows allows one to study the structures
formed by the total matter (dark and luminous) without the culprits of the redshift surveys which are probing only the bright galaxy distribution \citep{2014Natur.513...71T}.
Currently this field is dominated in number by galaxy distances derived using HI radio-detections for the Tully-Fisher relation \citep{2013AJ....146...86T}. However
the recent survey completed at the Schmidt telescope in Australia
\citep{2014MNRAS.443.1231C} is now adding a similar number of Fundamental Plane galaxy distance estimates. 

The second immediate use is for measurements of the local bulk flow and its physical interpretation for understanding the global motion towards the CMB dipole (Hoffman et al. 2015).
When various scales are tested, comparisons are possible with the observed velocity field computed from the redshift surveys distribution of galaxies \citep{2013IAUS..295..233M}.

Finally, a third use for such surveys is to provide even better constraints
on parameters of cosmological interest than a survey of redshifts alone \citep{2004MNRAS.347..255B,2005MNRAS.357..527Z}.
 Comparing the galaxy-galaxy, galaxy-velocity and velocity-velocity power spectra \citep{2014MNRAS.444.3926J}
  can lead to cosmological parameters, such as the redshift space distortion $\beta$ and the correlation between galaxies and dark matter r$_g$, 
that are degenerated when only the information provided by redshift surveys is used.

Those second generation surveys (LoRCA + TAIPAN) with high enough spectral resolution to measure velocity dispersions of early type galaxies, will allow us to probe larger scales than the ones reached today with the southern 6dFGSv sample, the northern SDSS FP sample and by single-dish Tully-Fisher (TF) peculiar velocity samples.
 
Together with new supernova surveys (e.g. \citealt{2012MNRAS.420..447T,2013MNRAS.431.3678R,2013MNRAS.428.2017M}) the new peculiar velocity surveys of thousands of galaxies will provide an unprecedented cartography of the motions in the local universe. Independent cross-checks and more accurate measurements, will complement the 3D density maps provided by redshift surveys and lead to tighter constraints on a wide range of cosmological parameters.

\subsection{Redshift space distortions}
The redshift of the galaxy is affected by the Hubble flow and by its peculiar velocity. The peculiar velocity makes the clustering anisotropic in redshift space \citep{1987MNRAS.227....1K}. This anisotropy is related to the growth rate of structure and to the amplitude of fluctuations, designated as $f\sigma_8$. With the LoRCA survey, one can measure $\xi(30 h^{-1}$ Mpc$)$ at the 6\% level, which should yield interesting constraints on the local growth rate of structure. We leave quantitative statements on the redshift space distortions for future studies.

\subsection{Tracing the mass function of galaxies down to low redshift}

The stellar mass function (SMF) of galaxies as a function of look-back
time, environment and galaxy type is one of the most important
statistics for studying galaxy formation and evolution on a
cosmological scale. While the SDSS-II used to set the scale for
the local ($z<0.1$) mass function (e.g. \citealt{2004ApJ...600..681B,2008MNRAS.388..945B,2009MNRAS.398.2177L}), an interesting discrepancy has emerged with
the higher redshift mass function calculated from SDSS-III
data ($z\sim0.5$, \citealt{2013MNRAS.435.2764M}). It is found that the
density of the most massive galaxies ($\log{(M/M_\odot)}>11.7$ for a Kroupa
IMF) at high-$z$, is higher than the correspondent in the local
Universe. 

This finding is intriguing as massive galaxies cannot
be disrupted. \citet{2013MNRAS.435.2764M} discuss possible causes for such
a mismatch, which include a smaller volume sampled locally
which would miss a fraction of the most massive galaxies or
systematics introduced by the different modeling of stellar
mass. \citet{2013MNRAS.436..697B} argue that in the local SDSS-based
SMF an incorrect evaluation of the light profile of massive
galaxies led to an underestimation of their light hence of their mass. 

Moreover, the comparison by \citet{2014ApJ...797L..27S} available mass functions up to $z\sim1$ showed that the \citet{2013MNRAS.436..697B} 
and the \citet{2013ApJ...773..112C} local mass functions are sizeably too
high compared to the $z\sim0.5$ mass functions from both surveys SDSS-III
\citep{2013MNRAS.435.2764M} and PRIMUS \citep{2013ApJ...767...50M}.

It seems then that the intermediate
redshift stellar mass function has been robustly assessed by 
previous surveys, while there is a clear call for a revision of the
local SMF. This is a goal we aim at achieving with the
LoRCA survey.

\subsection{Galaxy stellar populations}

The LoRCA survey set-up with about 8.6 arcsecond fibers will allow accurate studies of the stellar population constituting galaxies.

Using the CALIFA survey, that mapped spatially with integral field spectroscopy a sample of 400 local galaxies, \citep{2014A&A...562A..47G,2015A&A...581A.103G} demonstrated that the properties of the stellar population at the half light radius (HLR) are representative of:
\begin{enumerate}
\item the mean properties of the spatially resolved stellar populations of the galaxies;
\item the integrated properties of the galaxies, like the mean stellar age, the stellar mass density, metallicity and stellar extinction. 
\end{enumerate}

Using the values of redshift and half light radius provided in the catalog of \citet{2011AJ....142...31B}, we find that at z$\sim0.05$, 51\%\ of galaxies have an HLR$\leq$4 arc seconds and that at z$\sim0.15$, 96\%\ have their HLR$\leq$4 arc seconds.
Therefore, the spectroscopy obtained by LoRCA will provide enough spatial coverage to obtain representative integrated properties for the stellar populations for 75\% of the sample.

\subsection{Strong lensing}
Extrapolating from current strong lens sample (SLACS \citealt{2008ApJ...682..964B}), we expect to observe about 40 new strong lens system, which would double the current statistics and over twice the area currently covered.
These strong lens systems would be constituted of a massive elliptical $k<14$ lens at redshift $z<0.2$ and a strongly star forming galaxy at redshift $0.2<z<0.8$.

Such a sample would allow a better measurement of empirical scaling laws between radii, stellar and halo masses, colors and magnitude \citep{2009ApJ...705.1099A}. Such scaling laws are key to relating observations to the predictions of future cosmological simulations and in understanding the processes of galaxy formation.

\section{Summary}

Using the BigMD Planck simulation and EZmocks, we constructed a suite of a thousand light-cones to evaluate the precision with which one can measure the BAO standard ruler in the local universe.

We find that using the most massive galaxies drawn from 2MASS, $k$-selected to magnitude 14, available on the full-sky (34,000 deg$^2$), one can measure the BAO scale up to a precision of $\sim4\%$ and up to $\sim1\%$ after reconstruction. We also find that such a survey helps detect the dynamics of dark energy at $z<0.2$.

Therefore, it seems that the BAO standard ruler in the local universe is competitive with the predictions from the JWST Supernovae and more accurate than cepheids predictions. A non-negligible advantage of the BAO measurement compared to other methods is that the systematic errors are lower than the statistical errors.
We thus proposed a 3-year survey, Low Redshift Calar Alto (LoRCA) to build the largest local galaxy map possible that is easily implementable at low costs and for a very high scientific return.

All the light-cones, as well as their correlation functions and the template spectra are publicly available through the LoRCA website\footnote{\url{http://lorca-survey.ft.uam.es/lorca-mock-catalogs/}}.

%============= ============= ============= ========== 
%====================  Aknowledgements ===============
%=============== ===================  ===============

\section*{Acknowledgements}
This work is dedicated to the memory of Dr. Kurt Birkle\footnote{\url{http://www.caha.es/kurt-birkle-in-memoriam_es.html}}. The Schmidt telescope at Calar Alto was his passion, where he spent hundreds of observing nights over 25 years, since its official inauguration in 1979 at Calar Alto (moved from Hamburg Observatory, where it was commissioned in 1954); an event that marked the creation of the German-Spanish Astronomical Center (CAHA). His legacy is part of the HDAP - Heidelberg Digitized Astronomical Plates Project\footnote{\url{http://www.lsw.uni-heidelberg.de/projects/scanproject/}}, which digitized among others the unique collection of $5^\circ\times5^\circ$ photographic plates, as those of the Halley's comet in 1986; see the HDAP Database\footnote{\url{http://dc.zah.uni-heidelberg.de/lswscans/res/positions/q/form}}. 

We are grateful to David Schlegel for his valuable recommendations on the LoRCA instrument, and to the CAHA staff Jens Hemling, Santos Pedraz, and Jesus Aceituno for their help to provide archival documentation and key technical information on the Schmidt telescope. 

JC thanks J. Vega for insightful discussion about strong lensing. 

The authors wish to thank C. Blake, M. Colless and M. Bilicki for their insightful feedback on the draft.

JC, CC, SRT, MPI, FP, JS and FK acknowledge support from the Spanish MICINNs Consolider-Ingenio 2010 Programme under grant MultiDark CSD2009-00064, MINECO Centro de Excelencia Severo Ochoa Programme under grant SEV-2012-0249, and grant AYA2014-60641-C2- 1-P.
JC, GY acknowledges financial support from MINECO (Spain) under project number AYA2012-31101 and FPA2012-34694. MPI acknowledges support from MINECO under the grant AYA2012-39702-C02-01.
MPI thanks the Instituto de Fisica Teorica for its hospitality and support during the completion of this work
HC acknowledge support from the Lyon Institute of Origins under grant ANR-10-LABX-66.

GBZ is supported by the 1000 Young Talents program in China, and by the Strategic Priority Research Program "The Emergence of Cosmological Structures" of the Chinese Academy of Sciences, Grant No. XDB09000000.  

YW is supported by the National Science Foundation of China under Grant No. 11403034.

RBM's research is funded under the European Seventh Framework Programme, Ideas, Grant no. 259349 (GLENCO).

We acknowledge the work of Emilio Rodriguez, mechanical engineer at the IAA-CSIC, for the measurement of the Schmidt focal surface.

The BigMD simulation suite have been performed in the Supermuc supercomputer at LRZ using time granted by PRACE.

This work made use of the HDAP which was produced at Landessternwarte Heidelberg-Konigstuhl under grant No. 00.071.2005 of the Klaus-Tschira-Foundation.

\bibliographystyle{mn2e}
\bibliography{calar-alto-letter}

%============= ============= =============  ===========%=====================   Appendixes ====================
%=============== ===================  ================

\end{document}